\documentclass[aps,prd,singlecolumn,superscriptaddress,amsfont,graphicx,nofootinbib,preprintnumbers]{revtex4}
\pdfoutput=1
\usepackage{color,graphicx,epsfig}
\usepackage{ifpdf}
\usepackage{amsmath}
\usepackage{bm}
\usepackage{color}
\usepackage[english]{babel}
\usepackage{graphicx}
\usepackage{amsfonts}
\usepackage{amssymb}
\usepackage{braket}
\usepackage{hyperref}
\usepackage{multirow}

\definecolor{nicered}{rgb}{0.7,0.1,0.1}
\definecolor{nicegreen}{rgb}{0.1,0.5,0.1}
\hypersetup{colorlinks,citecolor= nicegreen,linkcolor= nicered}

\newcommand{\beq}{\begin{equation}}
\newcommand{\eeq}{\end{equation}}
\newcommand{\bea}{\begin{eqnarray}}
\newcommand{\eea}{\end{eqnarray}}

\definecolor{Red}{rgb}{1.,0.,0.}

\begin{document}

\def\LjubljanaFMF{Faculty of Mathematics and Physics, University of Ljubljana,
 Jadranska 19, 1000 Ljubljana, Slovenia }
\def\LjubljanaIJS{Jozef Stefan Institute, Jamova 39, 1000 Ljubljana, Slovenia}

%\preprint{}

\title{Light Higgs and Vector-like Quarks without Prejudice}

\author{Svjetlana Fajfer} 
\email[Electronic address:]{svjetlana.fajfer@ijs.si} 
\affiliation{\LjubljanaIJS}
\affiliation{\LjubljanaFMF}

\author{Admir Greljo} 
\email[Electronic address:]{admir.greljo@ijs.si} 
\affiliation{\LjubljanaIJS}

\author{Jernej F.\ Kamenik} 
\email[Electronic address:]{jernej.kamenik@ijs.si} 
\affiliation{\LjubljanaIJS}
\affiliation{\LjubljanaFMF}

\author{Ivana Musta\'c} 
\email[Electronic address:]{ivana.mustac@ijs.si} 
\affiliation{\LjubljanaIJS}

\date{\today}
\begin{abstract}
Light vector-like quarks with non-renormalizable couplings to the Higgs are a common feature of models trying to address the electroweak (EW) hierarchy problem by treating the Higgs as a pseudo-goldstone boson of a global (approximate) symmetry. We systematically investigate the implications of the leading dimension five operators on Higgs phenomenology in presence of dynamical up- and down-type weak singlet as well as weak doublet vector-like quarks. After taking into account constraints from precision EW and flavour observables we show that contrary to the renormalizable models, significant modifications of Higgs properties are still possible and could shed light on the role of vector-like quarks in solutions to the EW hierarchy problem. We also briefly discuss implications of higher dimensional operators for direct  vector-like quark searches at the LHC.
\end{abstract}

\maketitle

%%%%%%%%%%%%%%%%%%%%%%%%%%%%%%%%%%%%%%%%
%
\section{Introduction} 
%
%%%%%%%%%%%%%%%%%%%%%%%%%%%%%%%%%%%%%%%%

The recent LHC discovery of a Higgs boson~\cite{Aad:2012tfa} seems to  experimentally complete the standard model (SM) picture of fundamental interactions. Although the SM works extremely well phenomenologically,  the electro-weak (EW)  hierarchy problem, as exemplified by the extreme UV sensitivity of the Higgs potential, provides a strong theoretical motivation for contemplating beyond SM physics at the TeV scale. 

On the other hand, direct searches for on-shell production of new degrees of freedom at the LHC together with ever more stringent constraints from measurements of flavor, EW and Higgs observables are starting to directly probe models addressing the naturalness problem of the SM. At present these experimental null-results are not yet conclusive, but viable new physics (NP) models' parameter space is becoming significantly reduced. In light of this, it has become crucial to focus on the minimal light particle content required to fulfil the naturalness conditions (see~\cite{Papucci:2011wy} for examples in supersymmetric theories). 

Vector-like quarks are expected to be the lightest new degrees of freedom in models addressing the EW hierarchy problem by treating the light Higgs as a pseudo-goldstone boson of a global symmetry, broken explicitly by the SM gauging and Yukawa couplings~\cite{ArkaniHamed:2001nc,Panico:2012uw}. In such models, the dominant quadratic divergences in one-loop corrections to the Higgs boson mass coming from the top quark loop are canceled by contributions of dimension five operators of the form $H^{\dagger}H\bar{Q}Q$, where $H$ is the Higgs doublet and $Q$ a vector-like quark weak multiplet. In general, parametrizing single- and double-Higgs interactions of an arbitrary number of quark flavors $f$  in the mass eigenbasis
\beq
\mathcal L^{\rm eff}_{h} = \left(-y_{ij} h + x_{ij} \frac{h^2}{2v} \right) \overline{f}^i_L f^j_R + \rm h.c.\,,
\label{eq:LHiggs}
\eeq
where $v= (\sqrt 2 G_F)^{-1/2}\simeq 246$~GeV is the EW condensate, the condition for cancelation of one-loop quadratic divergences from such interactions can be put into the following simple form
\begin{equation}
\underset{i}{\sum}\Re(x_{ii})\frac{m^{}_{i}}{v}= \underset{i,j}{\sum}\left|y_{ij}\right|^{2}.
\label{eq:natural}
\end{equation} 
The appearance of new operator contributions already at dimension five is a singular feature of effective theories with new light vector-like fermions and intrinsically connected to the resolution of the SM hierarchy problem within such scenarios. These dimension five contributions are thus expected to represent the dominant probe of new dynamics in the UV.  At the same time one should keep in mind that unless the vector-like quarks are related to the SM field content by a symmetry, such cancelation of quadratic divergences is fine tuned in general. Furthermore, if one considers the running of the leading vector-like quark operator and the SM couplings, even if tuned to cancel at one scale, the quadratic divergences will not cancel at other scales. Thus, the effective theory discussion concerning  EW naturalness should be understood under the implicit assumption that the relation~\eqref{eq:natural} is enforced by symmetry in the UV complete theory. 

Furthermore,  even though a (symmetry enforced) relation~\eqref{eq:natural}  removes the quadratic UV sensitivity of the Higgs potential, logarithmically divergent contributions remain present in the effective theory. The biggest resulting shift to the bare Higgs mass ($\delta m_h$) is now due to the new heavy quark states with bilinear couplings to the Higgs. Assuming a single such state ($f$) cancelling the one-loop quadratically divergent contribution to $\delta m_h$ of the top quark, the dominant remaining correction is of the form
\beq
\delta m_h^2 \approx \frac{3 m_t^2}{4\pi^2 v^2} m_f^2 \log \frac{\Lambda^2}{m_f^2}\,,
\eeq
where the result was obtained using  a hard UV cut-off of the loop momentum integral and equating it with the cut-off scale of the effective theory $\Lambda$. We immediately observe that allowing for only moderate fine-tuning  (requiring conservatively $\delta m_h^2/m_h^2 \lesssim 10$) and the effective theory treatment valid (and thus $m_f \ll \Lambda$) requires $f$ to be relatively light ($m_f\lesssim 1$~TeV).

Treating the Higgs as a composite field of a strongly interacting theory leads to the appearance of a number of dimension six operators affecting flavor, EW and Higgs observables~(c.f.~\cite{Giudice:2007fh}). However, in models where the new vector-like quarks (possibly mixing with the chiral quark multiplets) and the Higgs are the lightest composite remnants of the strongly interacting sector, the dominant effects are expected from operators involving these fields. It is therefore meaningful to focus primarily on the leading dimension five contributions.\footnote{For a recent analysis of dimension six operator effects in composite Higgs scenarios without dynamical vector-like fermions see~\cite{Contino:2013kra}.} It turns out that they can always be parametrized in a way that preserves the form of gauge interactions of the renormalizable theory. Therefore the only way to approach and constrain such terms is by studying their impact on Higgs phenomenology, which is the main topic of the present work.\footnote{For recent related studies in the context of explicit composite Higgs model realisations see~\cite{Azatov:2011qy}.}  

The paper is structured as follows. In Sec.~\ref{sec:II} we present some general considerations of models with vector-like quarks  including generic flavor, electroweak and Higgs constraints on their interactions, stemming from renormalizable as well as leading dimension five non-renormalizable terms in the effective Lagrangian. Then in Secs.~\ref{sec:III} -- \ref{sec:V} we consider three specific model examples and analyze their viability in light of the derived direct and indirect constraints. For simplicity we will consider examples where vector-like quarks appear in existing weak representations of SM chiral quarks, thus allowing for kinetic mixing with chiral quark multiplets, with interesting phenomenological consequences. We summarize our conclusions in Sec.~\ref{sec:VI}. Several supporting derivations and analyses of experimental constraints have been relegated to the appendices.

%%%%%%%%%%%%%%%%%%%%%%%%%%%%%%%%%%%%%%%%
%
\section{General considerations}
\label{sec:II}
%
%%%%%%%%%%%%%%%%%%%%%%%%%%%%%%%%%%%%%%%%

%%%%%%%%%%%%%%%%%%%%%%%%%%%%%%%%%%%%%%%%
\subsection{Renormalizable models}
%%%%%%%%%%%%%%%%%%%%%%%%%%%%%%%%%%%%%%%%

Since all vector-like quark models under consideration only contain colored fermions in SM gauge representations and charges, we can start by considering the mass matrices of the up- and down-type quarks in the weak (chiral) eigenbasis
\beq
- \mathcal L_{\rm mass} = \bar u_L^i \mathcal M_u^{ij}  u_R^j + \bar d_L^i \mathcal M_d^{ij} d_R^j  + \rm h.c.\,,
\eeq
where the indices $i,j$ run over all dynamical quark flavors (including new vector-like generations).
The mass matrices $\mathcal M_{u,d}$ can be diagonalized via bi-unitary rotations as $\mathcal{M}_{u,d,\rm diag}=U_{L}^{u,d}\mathcal{M}_{u,d}U_{R}^{u,d\dagger}$\,. Consequently, the gauge and Higgs interactions of physical quarks in the mass eigenbasis can be written in the general  form (c.f.~\cite{AguilarSaavedra:2002kr})
\begin{align}
\label{eq:LW}
\mathcal L_W &= - \frac{g}{\sqrt 2} ( V^L_{ij} \bar u^i \gamma^\mu P_L  d^j + V^R_{ij}  \bar u^i \gamma^\mu  P_R d^j)  W^+_\mu   + \rm h.c.\,, \\
\label{eq:LZ}
\mathcal L_Z & = - \frac{g}{2 c_W} \left( X^u_{ij} \bar u^i \gamma^\mu P_L u^j - X^d_{ij} \bar d^i \gamma^\mu P_L d^j + Y^u_{ij} \bar u^i \gamma^\mu P_R u^j - Y^d_{ij} \bar d^i \gamma^\mu P_R d^j  - 2 s_W^2 J_{\rm EM}^\mu  \right) Z_\mu\,, \\
\label{eq:Lh0}
\mathcal L^{(0)}_h & = -   (X^u_{ij} - Y^u_{ij}) \frac{m_j}{v}   \bar u^i P_R u^j h - (X^d_{ij} - Y^d_{ij}) \frac{m_j}{v} \bar d^i P_R d^j h + \rm h.c.\,,
\end{align}
where $P_{R,L} = (1\pm\gamma_5)/2$, $g =2m_W/v \simeq 0.65$ is the weak coupling, while $s_W \simeq \sqrt{0.23}$ and $c_W = \sqrt{1-s_W^2}$ are the sine and cosine of the weak angle, respectively. $J^\mu_{\rm EM} = (2\bar u^i \gamma^\mu u^i -  \bar d^i\gamma^\mu d^i)/3$ is the EM quark current. The flavor matrices $V^{L,R}$, $X^{u,d}$ and $Y^{u,d}$ are all given in terms of $U^{u,d}_{L,R}$, in particular we can write $V^L_{ij} \equiv (U^d_{L})^*_{jk}  (U^u_{L})_{ik}$, where the repeated index runs over all left-handed weak doublets, and $V^R_{ij} \equiv (U^d_{R})^*_{jk}  (U^u_{R})_{ik}$, where the repeated index runs over all right-handed weak doublets. Then the (hermitian) flavor matrices entering neutral current  and Higgs interactions are given simply by $X^u \equiv V^L V^{L\dagger}$, $X^d \equiv V^{L\dagger} V^{L}$, $Y^u \equiv V^R V^{R\dagger}$ and $X^d \equiv V^{R\dagger} V^{R}$. Thus, non-standard Higgs interactions 
in such renormalizable models with extra quarks are necessarily constrained by charged and neutral weak currents among the known three generations of quarks.  For example, the departures of $V^L$ from $3\times 3$ unitary matrix and the appearance of a non-zero $V^R$ are constrained by precisely measured tree level charged current processes. For example, $\sum_{j=d,s,b} |V^L_{ij}|^2 = 1 - \Delta^u_{i} \leq 1$ (for $i=u,c,t$) and   $\sum_{j=u,c,t} |V^L_{ji}|^2 = 1 - \Delta^d_{i} \leq 1$ (for $i=d,s,b$) are constrained in absence of $V^R$  as $\Delta^u_{u} < 0.001$~\cite{Dowdall:2013rya}, $\Delta^u_{c} < 0.052$~\cite{pdg}, $\Delta^u_{t} < 0.13$ (see Appendix~\ref{sec:top} for details)\,, $ \Delta^d_{d} < 0.01$, $ \Delta^d_{s} < 0.08$~\cite{pdg} and $ \Delta^d_{b} < 1-|V^L_{tb}|^2<0.15$ (see Appendix~\ref{sec:top} for details). Note that in models with no extra up-type (down-type) quarks, $ \Delta^d_{i}  = \delta X^d_{ii}$ ($ \Delta^u_{i}= \delta X^u_{ii}$), where $\delta X^{u,d}_{ii} \equiv 1- X^{u,d}_{ii}$\,.
The entries of $V^R$ on the other hand, are also constrained at the tree-level by searches for right-handed charged currents (c.f.~\cite{Buras:2010pz} for a recent analysis). Unfortunately, without information on the matrix elements involving also extra quarks present in the model beyond the known SM generations, these cannot be directly related to $Z$ and Higgs couplings ($Y^{u,d}$)\,. 

In addition, one can obtain tree-level constraints on the off-diagonal entries of $X^{u,d}$ and $Y^{u,d}$ directly from their contributions to $Z$-mediated FCNCs of up- or down-type quarks. In all scenarios we consider, either nonstandard $X_{ij}^{u,d} \neq \delta_{ij}$ or $Y_{ij}^{ud} \neq 0$ are generated but not both. In this case the bounds on non-diagonal entries of $X^{u,d}$ or $Y^{u,d}$ read ${|X^u_{cu}|}, |Y^u_{cu}| < 2.1\times 10^{-4} $~\cite{Golowich:2007ka,hfag}, ${|X^u_{tu,tc}|,|Y^u_{tu,tc}|} < 0.14$ (see Appendix~\ref{sec:top} for details); ${\rm Re}(X^d_{ds}), {\rm Re}(Y^d_{ds})< 1.4\times 10^{-5}$, $|X^d_{db}|,|Y^d_{db}|< 4 \times 10^{-4}$ and $|X^d_{sb}|,|Y^d_{sb}|< 1\times 10^{-3}$ (see Appendix~\ref{sec:down} for details). Finally, electroweak measurements provide strong tree-level constraints also on the diagonal entries of $X^{u,d}$ and $Y^{u,d}$ corresponding to the five light quark flavours (see Appendix~\ref{sec:Z} for details). 

The main consequence of the above discussion is that in renormalizable models with additional vector-like quarks, Higgs couplings to the known three generations of quarks, except possibly the top (i.e. $X^u_{tt}$, $Y^u_{tt}$), must remain SM-like, irrespective of the spectrum or interactions of additional heavy quarks. This is because they are rigidly related to the corresponding $Z$ couplings, and thus subject to severe constraints from charged and neutral weak currents. %are insensitive to virtual (loop) contributions from new heavy states.  
In order to possibly obtain more interesting Higgs phenomenology, we are thus led to consider effects of higher-dimensional operators.

%%%%%%%%%%%%%%%%%%%%%%%%%%%%%%%%%%%%%%%%
\subsection{Including non-renormalizable Higgs interactions}
%%%%%%%%%%%%%%%%%%%%%%%%%%%%%%%%%%%%%%%%

Treating the SM as an effective field theory with particle content valid below a UV cut-off scale $\Lambda$, it is well known that  the leading higher dimensional operators involving quark fields are of dimension six, a virtue of the chiral nature of weak interactions in the SM. Thus, effects of NP degrees of freedom appearing above $\Lambda$ in low energy observables are suppressed by at least two powers of $1/\Lambda$. On the other hand, in presence of dynamical vector-like quarks, the leading non-renormalizable operators can appear already at dimension five. In general, they are of the form $H^\dagger H \bar Q Q$ and $H^\dagger H \bar q Q$, where $q$ denotes the SM chiral quark multiplets.\footnote{In the following we do not consider operators of the form $\bar Q (\sigma \cdot \mathcal G) Q$ and $\bar q (\sigma \cdot \mathcal G) Q$, where $\sigma_{\mu\nu} = i[\gamma_\mu,\gamma_\nu]/2$ and $\mathcal G_{\mu\nu} \in \{ T^a G^a_{\mu\nu}, \tau^a W^a_{\mu\nu}, B_{\mu\nu} \}$ stands for the three SM gauge field strengths, since these are not directly related to Higgs phenomenology. If the vector-like quarks mix with the SM generations, they will induce anomalous dipole gauge interactions of SM quarks at order $1/ m_Q \Lambda $, where $m_Q$ is the vector-like quark mass scale, and can be constrained from precision electroweak, flavor and collider observables (c.f.~\cite{dipole}). Modulo fine-tuned cancelations in these constraints, their presence would thus not affect our analysis.} 
The main consequences of these new interactions are (i) direct di-Higgs coupling to physical quarks [$x_{ij}$ in eq.~\eqref{eq:LHiggs}] with possible implications for the SM hierarchy problem; (ii) modifications of single Higgs - quark couplings  [$y_{ij}$ in eq.~\eqref{eq:LHiggs}] not related to weak neutral or charged currents. In the quark mass eigenbasis these additional contributions can generally be written as
\beq
\mathcal L_h^{(1)} = \left(\frac{X^{u\prime}_{ij}}{\Lambda} \bar u^i_L u_R^j  + \frac{X^{d\prime}_{ij}}{\Lambda} \bar d^i_L d_R^j \right) \left[ v h + \frac{h^2}{2}\right] + \rm h.c.\,,
\label{eq:LagHiggs1}
\eeq
where $\Lambda$ is the UV cut-off scale of the effective theory encompasing the SM together with a number of additional quark-like states. 
First note that the appearance of $X^{u,d\prime}_{ij}$ couplings of the known three generations of quarks to the Higgs is a manifestation of mixing between chiral and vector-like quarks, which is in general unrelated and thus unconstrained by charged and neutral weak currents. On the other hand, naturalness of the hierarchical quark mass spectrum would require $|X^{q\prime} _{ij} {X^{q\prime} _{ji}}^*|  v^4 /\Lambda^2  <  m_i m_j$\,~\cite{Cheng:1987rs}. In order to keep our analysis as general as possible, we shall not impose such a condition on the parameter space of our models, although one should keep it in mind.
The off-diagonal values of $X^{u,d\prime}$ are constrained by low energy flavour observables~\cite{Harnik:2012pb}. In the up-sector, $|X^{u\prime} _{uc,cu}|v/\Lambda<7\times10^{-5}$ and $\sqrt{|X^{u\prime} _{tu,tc}|^2 + |X^{u\prime} _{ut,ct}|^2}v/\Lambda<0.34$ are constrained by $D^0$ mixing and $t\to (c,u)h$ decay searches, respectively. Similarly, $K^0$, $B_d$ and $B_s$ mixing measurements require $|X^{d\prime}_{sd,ds}|v/\Lambda<2\times10^{-5}$, $|X^{d\prime}_{bd,db}|v/\Lambda<2\times10^{-4}$ and $|X^{d\prime}_{sb,bs}|v/\Lambda<1\times10^{-3}$, respectively.\footnote{All bounds from neutral meson mixing apply in absence of large cancellations with the tree-level $Z$-mediated $X^{u,d}, Y^{u,d}$ contributions.} 

Potentially the most striking tree-level effects on Higgs phenomenology in non-renormalizable vector-like quark models are thus modifications of the flavor diagonal Higgs couplings
to lighter quarks. In particular, in the SM the the total Higgs decay width is dominated by the $h\to b\bar b$ channel, the first hints of which have also been observed at the LHC~\cite{ATLAS-2013-034, CMS-12-045}. The modifications of $y_{bb}$ in eq.~\eqref{eq:LHiggs} can thus have important consequences for all experimentally observed Higgs signals. Similarly, while the $h\to u\bar u, d\bar d, s\bar s $ or $h\to c\bar c$ decays are very suppressed in the SM and also cannot be reconstructed at the LHC due to the large QCD backgrounds, they can contribute to the total Higgs decay width in case of non-zero $X^{u,d\prime}_{ii}$.
In particular, defining $\Delta \gamma \equiv   \sum_{f=d,u,s,c} \Gamma_{h\to f\bar f}/\Gamma_h^{\rm SM}$, we obtain $\sum_{i=d,s}{\left|X^{d\prime} _{ii}v/\Lambda -m_i/v \right|^{2}+ \sum_{i=u,c}\left|X^{u\prime} _{ii}v/\Lambda-m_{i}/v\right|^{2} } \simeq 10^{-3} {\Delta\gamma}$ showing that sizable enhancement in these decay channels is possible (although the required values of $X^{u,d\prime}_{ii}/\Lambda$ would necessarily violate the corresponding quark mass naturalness conditions) and that a non-trivial constraint on $X^{u,d\prime}_{ii}$ can in principle be obtained from the total Higgs decay width. This rises a question of importance of $\overline{u}u\to h$ (or to a lesser extent $\bar d d \to h$) production mechanism compared to the dominant $gg\to h$ mode at the LHC and Tevatron.  In the zero-width approximation and at leading order in QCD, the ratio of the relevant hadronic cross sections in the two cases can be written as
\begin{equation}
\frac{\sigma(p_1 p_2\to h)_{u\overline{u}}}{\sigma(p_1 p_2\to h)_{gg}}=\frac{\Gamma_{}(h\to\overline{u}u)}{\Gamma_{}(h\to gg)}\frac{\mathcal{L}_{p_1p_2}^{u\overline{u}}(\tau)}{\mathcal{L}_{p_1p_2}^{gg}(\tau)}\,,
\end{equation}
where $\tau\equiv m_{h}^{2}/s$ with $s$ being the invariant collider energy squared. The relevant luminosity functions at hadronic ($p_1 p_2$) colliders are given by
\begin{equation}
\mathcal{L}^{q_1q_2}_{p_1p_2} (x)=\frac{1}{1+\delta_{q_1 q_2}} \int_{x}^{1}\frac{dy}{y}[f^{p_1}_{q_1}(y)f^{p_2}_{q_2}(x/y) +  f^{p_2}_{q_1}(y)f^{p_1}_{q_2}(x/y) ] \,,
\end{equation}
where $f_{q_j}^{p_i}$ are the corresponding
parton distribution functions (pdfs). Using the LO MSTW2008~\cite{Martin:2009iq} set,
with the factorization and renormalization scales fixed to $m_{h}\simeq 125$~GeV, the ratio $\mathcal{L}_{pp}^{u\overline{u}}/\mathcal{L}_{pp}^{gg}$
for $\sqrt{s}=7\,\textrm{TeV}$ and $\sqrt{s}=14\,\textrm{TeV}$ is $3.9\%$ and $2.3\%$ respectively.\footnote{Using instead the NNLO MSTW2008 pdfs with the same scale choices yields ratios of  $4.0\%$ and $2.5\%$ respectively.} Thus, even assuming comparable
$h\to u\bar u$ and $h\to gg$ decay rates, the up-quark contribution to Higgs production at the LHC is below the $\sim 12\%$ theoretical uncertainties~\cite{Ball:2013bra}  of the dominant gluon fusion production cross section.
Conversely, at the Tevatron, we find the relevant ratio $\mathcal{L}_{p\bar p}^{u\overline{u}}/\mathcal{L}^{gg}_{p\bar p}$
for $\sqrt{s}=1.96\,\textrm{TeV}$ to be sizable $26\%$. However, given the low statistics in the gluon fusion Higgs production channel at the Tevatron~\cite{Group:2012zca}, this again gives no relevant constraint on $\Gamma (h\to\overline{u}u)$. In the future, enhanced di-Higgs production at the LHC could possibly offer a competitive constraint on $X^{u\prime} _{uu}/\Lambda$\,. In particular, the relevant LO hadronic cross section is given by
\beq
\sigma({p_1p_2\to hh})^{X^{\prime} } = \int_{4 \tau}^{1} dx\, \hat \sigma^{X^{\prime} }_{u\bar u \to hh} (x s) \mathcal{L}^{u\bar u}_{p_1p_2} (x) \,,
\eeq
where 
\beq
\hat{\sigma}_{u\bar{u}\to hh}^{X^{\prime}}(\hat{s})=\frac{|X_{uu}^{u\prime}|^{2}\beta_{h}}{64\pi\Lambda^{2}}\left(1+\frac{3m_{h}^{2}}{\hat{s}-m_{h}^{2}}\right)^{2}\,,
\eeq
and $\beta_h = \sqrt{1-4m_h^2/\hat s}$. Using the same pdf parameters as above we obtain $\sigma({pp\to hh})^{X^{\prime}} /[(|X^{u\prime}_{uu}|/0.03) (1{\rm TeV} / \Lambda) ]^2   \simeq 5 (11)$~fb at the $8$~TeV ($14$~TeV) LHC, compared to the SM LO predictions~\cite{Djouadi:1999rca} of $\sigma({pp\to hh})^{SM} = 4 (16)$~fb, respectively.

%%%%%%%%%%%%%%%%%%%%%%%%%%%%%%%%%%%%%%%%
\subsection{Impact of existing Higgs data}
%%%%%%%%%%%%%%%%%%%%%%%%%%%%%%%%%%%%%%%%

Most interesting effects involving light vector-like quarks in Higgs phenomenology appear at the one-loop level. In general, Higgs-fermion interactions of the form~\eqref{eq:LHiggs} will
contribute to gluon fusion production and  di-photon decay of the Higgs at one loop
\begin{align}
R_{gg}&\equiv\frac{\Gamma_{h\to gg}}{\Gamma_{h\to gg}^{SM}}\simeq\frac{\left|\underset{i}{\sum}y_{ii}\frac{v}{m_{i}}C(r_{i})F_{1/2}(\tau_{i})\right|^{2}}{\left|\frac{1}{2}F_{1/2}(\tau_{t})+\frac{1}{2}F_{1/2}(\tau_{b})\right|^{2}}\,, \\
R_{\gamma\gamma}&\equiv\frac{\Gamma_{h\to\gamma\gamma}}{\Gamma_{h\to\gamma\gamma}^{SM}}\simeq\frac{\left|F_{1}(\tau_{W})+\underset{i}{\sum}y_{ii}\frac{v}{m_{i}}d(r_{i})Q_{i}^{2}F_{1/2}(\tau_{i})\right|^{2}}{\left|F_{1}(\tau_{W})+\frac{4}{3}F_{1/2}(\tau_{t})\right|^{2}}\,,
\end{align}
where $d(r_{i})$ and $C(r_{i})$ are the dimension and index of the color
representation of $f_i$, respectively, and $Q_{i}$ is its electric
charge. The relevant loop functions $F_{1}(\tau)$ and $F_{1/2}(\tau)$ can be found e.g. in ref.~\cite{Djouadi:2005gi}, and $\tau_{i}\equiv m_{h}^{2}/(4m_{i}^{2})$.
In the limit of large fermion mass, $F_{1/2}(\tau)\to F_{1/2}(0)=4/3$. In the
SM, gluon fusion production, $gg\to h$, is dominated by the top quark
loop with $(1/2)F_{1/2}(\tau_{t})=0.688$, with minor contribution from the bottom quark $(1/2)F_{1/2}(\tau_{b})=-0.04+\imath0.06$.
Conversely, Higgs decay to two photons in the SM is dominated by the $W$ boson loop yielding
$F_{1}(\tau_{W})=-8.34$, and interfering destructively with the top quark
contribution of $(4/3)F_{1/2}(\tau_{t})=1.84$. It turns out that lighter quark contributions
to loop induced Higgs processes are negligible even if their couplings
to the Higgs saturate the limits from $\Delta \gamma$ as discussed below.

In order to evaluate the current constraints on modified Higgs interactions, we analyze the latest available Higgs data, presented in Table~\ref{tab:Data-used-in}. 
Our procedure follows closely those of similar previous analyses~\cite{Cacciapaglia:2012wb} and we refer the reader to those references for a detailed discussion of the current issues in using the data set. 
%\footnote{Formally, such effects are intertwined with our ignorance of higher order QCD effects in NP contributions relative to the SM, for which we assume the canonical size $\alpha_s(m_h)\sim 10\%$.} 

Measurements are given in terms of Higgs signal strengths normalized to SM predictions
\begin{equation}
\mu_{A\to h}^{ h\to B}=\frac{\sigma_{A\to h}}{\sigma_{A\to h}^{SM}}\frac{\mathcal B_{h\to B}}{\mathcal B_{h\to B}^{SM}}\,,
\end{equation}
where $A\to h$ and $h\to B$ stands for different production mode
and decay channel, respectively.  Experimental best-fit values and variances are denoted by $\hat{\mu}_{i}$ and $\hat{\sigma}_{i}^{2}$, respectively. Experimental collaborations generally provide plots with separate contribution from VBF plus VH, and from ggF plus ttH production channels for a given decay channel. In this case, we take into account their correlation, obtaining a correlation parameter $\rho$ by reproducing the plots. The contribution from these decay channels to the total $\chi^{2}$ function is
\begin{equation}
\chi_{1}^{2}=\underset{i}{\sum}\left(\begin{array}{cc}
\mu_{GF}^{i}-\hat{\mu}_{GF}^{i}, & \mu_{VF}^{i}-\hat{\mu}_{VF}^{i}\end{array}\right)\left(\begin{array}{cc}
\left(\hat{\sigma}_{GF}^{i}\right)^{2} & \rho^{i}\hat{\sigma}_{GF}^{i}\hat{\sigma}_{VF}^{i}\\
\rho^{i}\hat{\sigma}_{GF}^{i}\hat{\sigma}_{VF}^{i} & \left(\hat{\sigma}_{VF}^{i}\right)^{2}
\end{array}\right)^{-1}\left(\begin{array}{c}
\mu_{GF}^{i}-\hat{\mu}_{GF}^{i}\\
\mu_{VF}^{i}-\hat{\mu}_{VF}^{i}
\end{array}\right),
\end{equation}
where $GF$ stands for ggF+ttH, and $VF$ stands for VBF+VH, and index $i$ runs over decay channels.\footnote{ttH contribution in $GF$ is less then $1\%$.} If the separation into production modes is not provided, we use the data from different search categories for a particular decay channel,  which generally target certain production mechanism, but does not imply $100\%$ purity. Inclusive categories are dominated by ggF ($\thicksim90\%$), while VBF-tagged categories can have $20\%$ to $50\%$ contamination from ggF.  VH- and ttH-tagged categories are assumed to be pure. In this case, we write
\begin{equation}
\frac{\sigma_{A\to h}}{\sigma_{A\to h}^{SM}}=\xi_{ggF}\frac{\sigma_{ggF}}{\sigma_{ggF}^{SM}}+\xi_{VBF}\frac{\sigma_{VBF}}{\sigma_{VBF}^{SM}}+\xi_{VH}\frac{\sigma_{VH}}{\sigma_{VH}^{SM}}+\xi_{ttH}\frac{\sigma_{ttH}}{\sigma_{ttH}^{SM}},
\end{equation}
where $\xi_{i}$ represent contributions of the specified production mechanisms for the given category. We do not assume correlations here, and add each search category to $\chi^{2}$ separately,
\begin{equation}
\chi_{2}^{2}=\underset{j}{\sum}\left(\frac{\mu_{j}-\hat{\mu}_{j}}{\hat{\sigma}_{j}}\right)^{2},
\end{equation}
and the total $\chi	^{2}$ function is given by $\chi^{2}=\chi_{1}^{2}+\chi_{2}^{2}$. 

We take into account the recent evaluation~\cite{Ball:2013bra} of ggF production in the SM at approximate N$^{3}$LO in perturbative expansion, which exhibits  a $17\%$ shift with respect to the values adopted by experimental collaborations, by rescaling central values for signal strengths which depend on ggF production by a factor $1/(1+0.17\xi_{ggF})$. On the other hand, we have checked that the resulting slightly reduced theory error (from $14\%$ to $12\%$) has negligible effect on the reported variances. Finally, we note that using $7$~TeV and $8$~TeV data combinations introduces potentially non-negligible effects due to the different parton luminosities when constraining NP. We expect however these to be subleading compared to the overall theoretical uncertainty, especially in light of our ignorance of higher order QCD effects in NP contributions.

\begin{table}
\begin{centering}
\begin{tabular}{|c|c|c|c|}
\hline 
Decay channel & Production mode & Signal strength & Comment\tabularnewline
\hline\hline 
\multicolumn{4}{|c|}{ATLAS}\tabularnewline
\hline\hline 
$h\to ZZ^{*}$ & Inclusive ($87\%\,$ggF) & $1.5\pm0.4$ & \cite{key-1,ATLAS-2013-034}\tabularnewline
\hline 
$h\to b\overline{b}$ & VH & $-0.4\pm1.0$ & \cite{ATLAS-2013-034}\tabularnewline
\hline 
\multirow{2}{*}{$h\to WW^{*}$} & ggF+ttH & $0.79\pm0.35$ & \multirow{2}{*}{Correlation $\rho=-0.3$, \cite{ATLAS-2013-034,key-W}}\tabularnewline
\cline{2-3} 
 & VBF+VH & $1.6\pm0.8$ & \tabularnewline
\hline 
\multirow{2}{*}{$h\to\gamma\gamma$} & ggF+ttH & $1.60\pm0.44$ & \multirow{2}{*}{Correlation $\rho=-0.4$, \cite{key-3,ATLAS-2013-034}}\tabularnewline
\cline{2-3} 
 & VBF+VH & $1.80\pm0.87$ & \tabularnewline
\hline 
\multirow{2}{*}{$h\to\tau\tau$} & ggF+ttH & $2.2\pm1.6$ & \multirow{2}{*}{Correlation $\rho=-0.5$, \cite{ATLAS-2013-034}}\tabularnewline
\cline{2-3} 
 & VBF+VH & $-0.3\pm1.1$ & \tabularnewline
\hline\hline 
\multicolumn{4}{|c|}{CMS}\tabularnewline
\hline\hline 
$h\to b\overline{b}$ & VH & $1.3\pm0.7$ & \cite{CMS-12-045}\tabularnewline
\hline 
\multirow{3}{*}{$h\to WW^{*}$} & 0/1 jet ($97\%\,$ggF) & $0.76\pm0.21$ & \cite{key-Cw}\tabularnewline
\cline{2-4} 
 & VBF-tag ($20\%\,$ggF) & $0.0\pm0.7$ & \cite{CMS-12-045}\tabularnewline
\cline{2-4} 
 & VH & $-0.3\pm2.1$ & \cite{CMS-12-045}\tabularnewline
\hline 
\multirow{2}{*}{$h\to ZZ^{*}$} & ggF+ttH & $0.90\pm0.45$ & \multirow{2}{*}{Correlation $\rho=-0.7$, \cite{key-5}}\tabularnewline
\cline{2-3} 
 & VBF+VH & $1.0\pm2.3$ & \tabularnewline
\hline 
\multirow{2}{*}{$h\to\gamma\gamma$} & ggF+ttH & $0.52\pm0.40$ & \multirow{2}{*}{Correlation, $\rho=-0.5$, \cite{key-9}}\tabularnewline
\cline{2-3} 
 & VBF+VH & $1.5\pm0.9$ & \tabularnewline
\hline 
\multirow{3}{*}{$h\to\tau\tau$} & 0/1 jet ($80\%\,$ggF) & $0.73\pm0.51$ & \cite{key-tau}\tabularnewline
\cline{2-4} 
 & VBF-tag ($20\%\,$ggF) & $1.37\pm0.63$ & \cite{key-tau}\tabularnewline
\cline{2-4} 
 & VH & $0.75\pm1.5$ & \cite{key-tau}\tabularnewline
\hline 
\end{tabular}
\par\end{centering}
\caption{LHC Higgs data used in the analysis. \label{tab:Data-used-in}}
\end{table}

In the following, we present results of our analysis. The SM is in overall very good agreement with the data. The associated $\chi^{2}$ for $19$ observables presented in Table~\ref{tab:Data-used-in} is $\chi_{SM}^{2}=16.2$ corresponding to a p-value of $0.64$. Within the vector-like quark scenarios, all the modifications to Higgs signal strengths can be expressed in terms of four parameters, $R_{gg}$, $R_{\gamma \gamma}$, $R_{bb}$ and $\Delta \gamma$, where
\begin{equation}
R_{bb}\equiv\frac{\Gamma_{h\to bb}}{\Gamma_{h\to bb}^{SM}}=\left( \frac{|y_{bb}|v}{m_b}\right)^2.
\label{eq:htobb}
\end{equation}
In particular, one can write
\begin{align}
\mu_{GF}^{h\to\gamma\gamma}&=\frac{R_{gg}}{\hat{\Gamma}}R_{\gamma\gamma}\,, & \mu_{GF}^{h\to ZZ,WW,\tau \tau}&=\frac{R_{gg}}{\hat{\Gamma}}\,, & \mu_{VF}^{h\to\gamma\gamma}&=\frac{R_{\gamma\gamma}}{\hat{\Gamma}}\, & \mu_{VF}^ {h\to ZZ,WW,\tau \tau} & =\frac{1}{\hat{\Gamma}}\,, & \mu_{VH}^ {h\to\overline{b}b} & =\frac{R_{bb}}{\hat{\Gamma}}.
\end{align}
The modification of the total Higgs decay width coming from $R_{gg}$, $R_{bb}$ and $\Delta\gamma$ is taken into account by writing
\begin{equation}
\hat{\Gamma}\equiv\frac{\Gamma_{tot}}{\Gamma_{tot}^{SM}}=0.569R_{bb}+0.317+0.085R_{gg}+\Delta\gamma\,,
\label{eq:htosve}
\end{equation}
where $\Delta \gamma$ is constrained to $\Delta \gamma >0$.

We consider four different scenarios, with different choices of the fitting parameters. In each case, the best-fit point is determined by minimizing the $\chi^{2}$ function. Results are presented in the $(R_{gg},R_{\gamma \gamma})$ plane, after marginalizing over the other parameters.  We define $68.2$\%
($1\sigma$) best-fit region to satisfy $\chi_{min}^{2}<\chi^{2}<\chi_{min}^{2}+2.3$, and $95.5$\% (2$\sigma$) best-fit region to satisfy $\chi_{min}^{2}+2.3<\chi^{2}<\chi_{min}^{2}+6.2$. 

\begin{figure}
\begin{centering}
\includegraphics[scale=1.5]{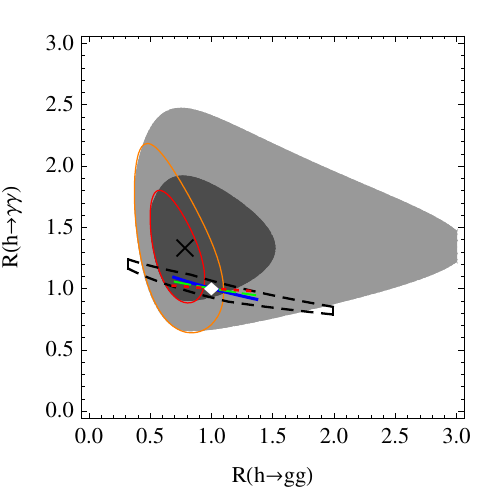} 
\includegraphics[scale=1.5]{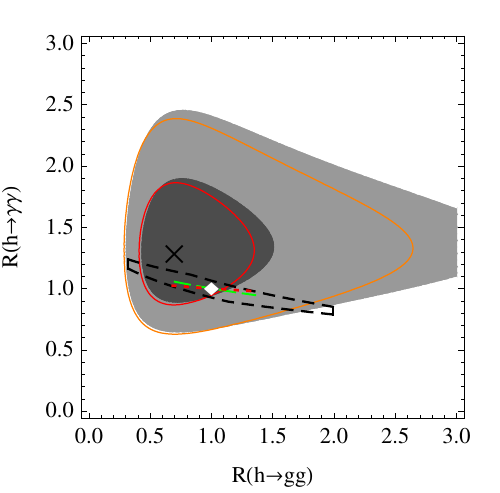} 
\par\end{centering}
\caption{\textbf{Left:} Fit of Higgs data taking $R_{gg}\equiv R(h\to gg)$, $R_{\gamma \gamma}\equiv R(h\to \gamma \gamma)$ and $\Delta \gamma$ as fitting parameters. The best fit point (cross), $1\,\sigma$ (dark gray) and $2\,\sigma$ (light gray) regions are shown in the $(R_{gg},R_{\gamma \gamma})$ plane after marginalizing over $\Delta \gamma$. The SM reference scenario is marked with a diamond. Results of the fit for $\Delta \gamma$ fixed to its SM value are given by the darker red contour ($1\,\sigma$ region) and the lighter orange contour ($2\,\sigma$ region). The resulting prediction from the non-renormalizable model with an additional up-like vector-like quark, the model with a down-like vector-like and the model with a doublet vector-like quark in the no-mixing and negligible isospin breaking limit are given by the continuous-blue, dotted-red and dashed-green curves, respectively. General predictions from the non-renormalizable model with a doublet vector-like quark are given by the region presented with the dashed-black contour (see text for details).
\textbf{Right:} Fit of Higgs data taking $R_{gg}$, $R_{\gamma \gamma}$, $R_b$ and $\Delta \gamma$ as fitting parameters. The best fit point (cross), $1\,\sigma$ (dark gray) and $2\,\sigma$ (light gray) regions are shown in the $(R_{gg},R_{\gamma \gamma})$ plane after marginalizing over $R_b$ and $\Delta \gamma$. Results of the fit for $\Delta \gamma$ fixed to its SM value are given by the darker red contour ($1\,\sigma$ region) and the lighter orange contour ($2\,\sigma$ region). The resulting prediction from the non-renormalizable model with a down-like vector-like and the model with a doublet vector-like quark in the no-mixing and negligible isospin breaking limit are given by dotted-red and dashed-green curves, respectively. General predictions from the non-renormalizable model with a doublet vector-like quark are given by the region presented with the dashed-black contour  (see text for details).
\label{fig:-plot-1}}
\end{figure}

First, we take $R_{gg}$ and $R_{\gamma \gamma}$ as fitting parameters, while fixing $R_{bb}$ and $\Delta \gamma$ to their SM values. This applies to scenarios, where the couplings of SM quarks to the Higgs (the Yukawas) are not modified, while $R_{gg}$ and $R_{\gamma \gamma}$ receive new loop contributions. In models with vector-like quarks this corresponds to the limit of zero mixing between the chiral and vector-like quarks. The minimum of the $\chi^{2}$ corresponds to a point $(R_{gg},R_{\gamma \gamma})=(0.71,1.30)$ with $\chi_{min}^{2}=12.7$ and p-value $0.76$.
Second, we take $R_{gg}$, $R_{\gamma \gamma}$ and $\Delta \gamma$ as fitting parameters, while fixing $R_{bb}=1$. This corresponds to scenarios where the vector-like quarks do not mix with the $b$ quark, but possibly with the lighter quark generations (see~\cite{Delaunay:2013iia} for a recent model example).  In this case, the point  $(R_{gg},R_{\gamma \gamma},\Delta \gamma)=(0.79,1.33,0.12)$ corresponds to the minimum of the $\chi^{2}$, with $\chi_{min}^{2}=12.6$ and p-value $0.7$. The results for the first two scenarios are presented on the left plot in Fig.~\ref{fig:-plot-1}. For the first scenario, $1\,\sigma$ and $2\,\sigma$ contours are represented by (darker) red and (lighter) orange curves, respectively.

In the third case, we take $R_{gg}$, $R_{\gamma \gamma}$ and  $R_{bb}$  as fitting parameters, while fixing $\Delta \gamma$ to its SM value.  This case applies to the most studied scenarios in the literature, where the vector-like quarks only mix with the third generation (c.f.~\cite{Kearney:2012zi} for a recent study). The minimum of the $\chi^{2}$ corresponds to a point $(R_{gg},R_{\gamma \gamma},R_{bb})=(0.70,1.29,0.97)$ with $\chi_{min}^{2}=12.7$ and p-value $0.69$. In the last case, we take all four parameters to fit the data, corresponding to the most general case of vector-like quarks mixing with all three SM generations. Here, the minimum of the $\chi^{2}$ corresponds to a point $(R_{gg},R_{\gamma \gamma},R_{bb},\Delta \gamma)=(0.73,1.31,0.74,0.2)$ with $\chi_{min}^{2}=12.4$ and p-value $0.65$. The results for the last two scenarios are presented on the same plot, Fig.~\ref{fig:-plot-1} right. For the third scenario, $1\,\sigma$ and $2\,\sigma$ contours are represented by (darker) red and (lighter) orange curves, respectively.

The main observation at this point is that allowing for a modification of $R_{bb}$ and/or $\Delta \gamma$ (non-zero mixing of vector-like quarks with some of the SM chiral quarks) significantly increases the allowed range of $R_{gg}$, while it has much less of an effect on $R_{\gamma\gamma}$\,. This has important implications for constraining vector-like quark models using Higgs data, since these generically predict much larger effects in $R_{gg}$. In the following sections we apply these general results to a few simplest SM extensions with a single light vector-like quark state below the effective theory cut-off $\Lambda$.

%%%%%%%%%%%%%%%%%%%%%%%%%%%%%%%%%%%%%%%%
%
\section{Singlet up-type vector-like quark}
\label{sec:III}
%
%%%%%%%%%%%%%%%%%%%%%%%%%%%%%%%%%%%%%%%%

%%%%%%%%%%%%%%%%%%%%%%%%%%%%%%%%%%%%%%%%
\subsection{Renormalizable model}
\label{sec:IIIa}
%%%%%%%%%%%%%%%%%%%%%%%%%%%%%%%%%%%%%%%%

As a first example, we consider the SM extended by a vector-like quark pair $(U_{L},U_{R})$
in the $\mathbf{1}_{2/3}$ representation of the SM electroweak group. In the most general renormalizable model the  quark Yukawa interactions and mass terms can be described by the following Lagrangian 
\begin{equation}
-\mathcal{\mathcal{L}}^{(0)}_{U}= y^{ij}_{d}{\bar q^i_{L}}{H}d^j_{R} + y^{ij}_{u}{\bar q^i_{L}}\tilde{H}u^j_{R}+y^i_{U}{\bar q^i_{L}}\tilde{H}U_{R}+M_{U}{\bar U_{L}}U_{R}+\textrm{h.c.}\,,\label{eq:Lagtop}
\end{equation}
where $\tilde H \equiv i \tau_2 H^*$, $H = (G^+,(v+h+ i G_0)/\sqrt 2)$ is the SM Higgs doublet, $q^i_{L}$ the SM quark doublets and $u^i_{R}$ the SM up-type quark singlets. Note that additional kinetic mixing terms of the form $\overline{U_{L}}u^i_{R}$ can always be rotated away and reabsorbed into the definitions of $y_{u,U}$. Furthermore, one can, without loss of generality, choose a weak interaction basis where $y_u$ is diagonal and real.  After EW symmetry breaking (EWSB) the mass matrices for up- and down-type quarks are
\beq 
\mathcal{M}_{u} = 
\left(\begin{array}{cc}
 y_u v/\sqrt 2\, & y_U v/\sqrt 2 \\
 0\, & M_U
\end{array} \right)\,, \quad \mathcal M_d = (y_d v /\sqrt 2) \,.
\label{eq:Mu}
\eeq
The weak gauge and Higgs interactions of 4 ($u,c,t,u'$) physical up-like and 3 ($d,s,b$) down-like quarks in this (mass) eigenbasis are given by eqs.~\eqref{eq:LW}-\eqref{eq:Lh0}, where $V^R=0$, $V^L$ is a general $4\times 3$ matrix and $X^d = \mathbb{ I}_{3\times 3}$\,. Note that in this model $X^u_{ii} = 1 -\Delta^u_{i}$ and that tree level constraints on the entries of $X^u$ already severely constrain the admixture of $U$ within the physical $u$ and $c$ quarks. In particular, we find for the $3\times 3$ sub-matrix of $X^u$ describing $Z$ and Higgs couplings to known up-type quark flavours
\beq
|X^{u}-\mathbb I |_{3\times 3} < \left[ \begin{array}{ccc}
0.001 & 2.1 \times 10^{-4} & 0.14  \\
  & 0.0026 & 0.14 \\
  & & 0.13
 \end{array} \right]\,.
\eeq
Loop-level $u'$ effects provide better constraints only on the mixing of the vector-like singlet quark with the top quark.  Neglecting the small mixing with the first two generations (effectively setting $y_U^{u,c}=0$) the $t-u'$ system can be described by three independent physical parameters: two quark masses ($m_t, m_{u'}$) and a single (left-handed) mixing angle ($\theta_{tU}$), which are defined as~\cite{Sally}
\begin{align} 
\tan (2\theta_{tU}) & = \frac{\sqrt 2 v y_U^t M_U}{M_U^2 - [(y_u^{t})^2 + (y_U^{t})^2]v^2/2}\,, \\
m_t m_{u'} &=  M_U y_u^t \frac{v}{\sqrt 2}\,, \qquad m_t^2 + m_{u'}^2 = M_U^2 + \frac{v^2}{2} [(y_u^{t})^2 + (y_U^{t})^2]\,.
\end{align}
In terms of these, $X^u_{tt} = c^2_{tU}$, $X^u_{tu'} = c_{tU} s_{tU}$ and $X^u_{u' u'} = s^2_{tU}$\,, where $c_{tU} \equiv \cos \theta_{tU}$ and $s_{tU} \equiv \sin \theta_{tU}$.

\begin{figure}
\begin{centering}
\includegraphics[scale=1.1]{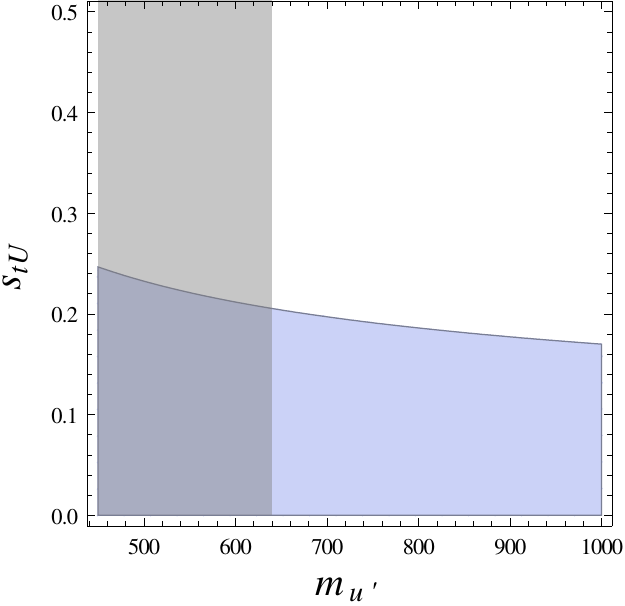}
\end{centering}
\caption{Upper limit at $95\%$ C.L. on $t - u'$ (left-handed) mixing angle as a function of the $u'$ quark mass in the model with an up-like vector-like quark. The gray region marks the ATLAS experimental search bound on the renormalizable model using the $u' \to t h$ decay signature \cite{atlas-bound}.}
\label{fig:1}
\end{figure}

Presently, the most sensitive observable to nonzero $s_{tU}$ is the $\rho$ parameter, which receives a new contribution of the form~\cite{Sally}
\beq
\Delta \rho = \frac{\alpha N_C }{16\pi s_W^2} \frac{m_t^2}{m_W^2} s^2_{tU} \left[  -(1+c^2_{tU})  + s^2_{tU} r + 2 c^2_{tU} \frac{r}{r-1}\log (r) \right]\,,
\eeq
where $r\equiv m_{u'}^2/m_t^2$ and we have neglected terms of higher order in $m_{Z,b}^2/m_{t,u'}^2$\,. A comparison with the experimental bound of $\Delta \rho^{\rm exp} = 4^{+3}_{-4} \times 10^{-4}$~\cite{pdg} yields a constraint on $s_{tU}$ as a function of the $u'$ mass as shown in Fig.~\ref{fig:1}.

While the modified top quark coupling to the Higgs boson
and the presence of an additional heavy quark can in principle impact also loop induced
Higgs decays, namely $h\to gg$, $h\to\gamma\gamma$ and $h\to Z\gamma$,
 taking into account the above constraints on $X^u_{ij}$ these effects turn out severely suppressed making it  impossible in practice to distinguish the renormalizable model with a singlet vector-like up type quark from the SM in single Higgs production processes.

%%%%%%%%%%%%%%%%%%%%%%%%%%%%%%%%%%%%%%%%
\subsection{Non-renormalizable models}
\label{sec:IIIb}
%%%%%%%%%%%%%%%%%%%%%%%%%%%%%%%%%%%%%%%%

Extending the above renormalizible model with the leading dimension five operators containing the light SM fields and $U_{L,R}$ as the only dynamical degrees of freedom below a UV cut-off scale $\Lambda$, Yukawa interactions and mass terms in eq.~\eqref{eq:Lagtop} receive corrections which result in modified interactions between up-type quarks and the Higgs. One can manifestly preserve the exact mass diagonalization procedure of the renormalizable model by parametrizing the leading non-renormalizable contributions in terms of the replacement
\begin{align}
M_{U} &\to M_U +  c_2 \frac{v^2/2-|H|^2}{\Lambda}\,, 
\end{align}
plus an additional Higgs-dependent `kinetic mixing' operator
\beq
- \mathcal L_U^{(1)} = c^i_1 \frac{v^2/2-|H|^2}{\Lambda} \bar U_L u^i_R\,.
\eeq 
After EWSB, the flavor structure of gauge interactions (and the associated bounds on $X^u_{ij}$) in the renormalizable model is preserved and only the Higgs interactions in the mass eigenbasis receive new contributions of the form~\eqref{eq:LagHiggs1} where $X^{d\prime} = 0$ (leading to $R_{bb}=1$) while $X^{u\prime} = U_L^u . [(0,0),(c_1  ,c_2 )] . U_R^{u\dagger}$.  Interestingly, even though $X^{u\prime} $ has no observable effects on charged current interactions of quarks, one can derive an indirect bound on the diagonal entries of $X^{u\prime} $ from CKM unitarity. Following its definition, $X^{u\prime} _{ij}=(U_{L}^{u})_{i4}\left(c_{1}^{k}(U_{R}^{u})_{jk}^{*}+c_{2}(U_{R}^{u})_{j4}^{*}\right)$, we note that $|X^{u\prime} _{ij}|^2$ is proportional to 
$|(U_{L}^{u})_{i4}|^2=1-X^u_{ii}$, multiplied by at most $\mathcal O(1)$ coefficients in the effective field theory expansion $c^i_{1},c_2$ and the unitary rotation $U^u_R$. Furthermore, the general identity $|(U^u_R)_{i4}|^2 = (m_i/M_U)^2 |(U_{L}^{u})_{i4}|^2$ implies that $c_2$ contributions to $u,c$ interactions are severely suppressed. Consequently, CKM unitarity constraints on $X^u_{uu,cc}$ yield  indirect bounds on the diagonal elements $\left|X^{u\prime}_{uu}\right|\lesssim0.03\, {\rm max} (c^i_1)$ and $\left|X^{u\prime}_{cc}\right|\lesssim0.2 \,{\rm max} (c^i_1)$.  Note that sizable contributions to $\Delta\gamma$ are well consistent with these indirect CKM unitarity constraints. 

The relevant modifications to $R_{gg}$ and $R_{\gamma\gamma}$ in the up-type singlet scenario come from the modified top quark coupling and the presence of the additional heavy quark in the loop. In the limit $m_{t,u'}\gg m_h$, which is a good approximation here, the relevant numerical expressions are given by
\begin{align}
R_{gg}&=\frac{\left|0.68 r_y -0.040\right|^{2}+\left.0.057\right.^{2}}{0.65^{2}}\,,\quad
R_{\gamma\gamma}=\frac{\left|-8.3+1.8r_y\right|^{2}}{\left|-6.5\right|^{2}}\,,
\end{align}
where to order $1/\Lambda$ and including possible small $t-u'$ mixing
  \begin{equation}
 r_y \equiv y_{tt}\frac{v}{m_{t}}+y_{u'u'}\frac{v}{m_{u'}}= 1+s_{tU}\left(c_{1}^{t}c'_{tU}-c_{2}s'_{tU}\right)\frac{v^{2}}{\Lambda m_{t}}-c_{tU}\left(c_{1}^{t}s'_{tU}+c_{2}c'_{tU}\right)\frac{v^{2}}{\Lambda m_{u'}}\,.
 \end{equation}
Above we have used the short-hand notation for $c'_{tU}=\cos \theta'_{tU}$, $s'_{tU}=\sin \theta'_{tU}$, where the (right-handed) mixing angle $\theta'_{tU}$ is defined via $\tan\theta'_{tU}=(m_{t}/m_{u'})\tan\theta_{tU}$. Thus, $h\to\gamma\gamma$ and $h\to gg$ are highly correlated in this set-up. Note that at the one loop level and in the large $m_{t,u'}$ limit, contributions of renormalizable interactions of $t$ and $u'$ cancel exactly\footnote{Deviations from the large mass limit, as well as higher order perturbative corrections can upset this cancellation. However a recent study of gluon fusion production in the renormalizable model with a singlet vector-like top partner at NNLO in QCD~\cite{Sally} has found such effects to be tiny, only a few percent for maximal mixing.}, therefore, leading effects appear  at order $\mathcal{O}(v/\Lambda)$. 

The resulting predictions in the up-type singlet scenario are presented by the continuous (blue) curve
in the $(R_{gg},R_{\gamma\gamma})$ plain in left plot of Fig.~\ref{fig:-plot-1}. For concreteness we take $\left|(c_{1}^{t},c_{2})\right|/\Lambda \leq1$~TeV${}^{-1}$ and $m_{u'}\geq640$~GeV as suggested by direct searches~\cite{atlas-bound} (see also the related discussion at the end of this section). Also, we take the $t-u'$ mixing angle to be within the $95\%$ C.L. experimental bound discussed above ($s_{tU}\lesssim 0.2$). From the fit to Higgs data we obtain the preferred parameter ranges for $r_y$ at $68\%$ C.L. of  $r_{y}=0.86_{-0.09}^{+0.16}$ (when marginalizing over $\Delta \gamma$) and $r_{y}=0.87\pm0.08$ (when fixing $\Delta \gamma=0$). Interestingly, in both cases the fit slightly prefers $r_y<1$. We note in passing that after marginalising over $r_{y}$  in this scenario $\Delta \gamma$ is bounded as $\Delta\gamma<0.75$ at $95\%$ C.L.\,, which in term implies $|X_{uu}^{u\prime}+X_{cc}^{u\prime}|{v}/{\Lambda}<0.022$\,.

Finally, turning to the naturalness condition in~\eqref{eq:natural}, for the case of a single vector-like up-type singlet quark mixing with the top it reads
\begin{equation}
\frac{m_{t}^{2}c_{tU}^2 + m^2_{u'} s_{tU}^2 }{v^{2}} = \frac{1}{\Lambda}[ m_t s_{tU}(  - c_1^t c'_{tU} + c_{2}s'_{tU} ) + m_{u'} c_{tU}   ( c_1^t  s'_{tU} + c_{2} c'_{tU}  )] + \mathcal O(1/\Lambda^2).\,
\end{equation}
In the zero-mixing limit  this leads to the prediction $r_y = 1-(m_t/m_{u'})^2$, independent of the cut-off scale $\Lambda$. Interestingly, present Higgs data (exhibiting a preference for $r_y<1$) are perfectly consistent with the naturalness condition. On the other hand,  the Higgs fit results can also be interpreted in this context as imposing an indirect bound on the $u'$ mass of  $m_{u'}>360$~GeV at 95\% C.L.\footnote{For a comparison with the situation after the first Higgs data see~\cite{Carmi:2012yp}.}

It is instructive to compare the above constraint to results of direct experimental searches for up-type singlet vector-like quarks. Interestingly, the most severe bound on $u'$ in the renormalizable model and assuming dominant but small $u'$ mixing with the top, $m_{u'}>640$~GeV~\cite{atlas-bound} is given by the ATLAS experimental search using the $u' \to t h$ decay signature. In the non-renormalizable model the relevant couplings are given by $y_{tu'} =  s_{tU} c_{tU} m_{u'}/v +s_{tU}  (s'_{tU} c_1^t + c'_{tU}c_2 )v/\Lambda$ and $y_{u't} =  s_{tU} c_{tU} m_{t}/v - c_{tU}  (c'_{tU} c_1^t -  s'_{tU}c_2)v/\Lambda$. It is then easy to check that compared to $u' \to t Z$ and $u' \to b W$ rates, the $1/\Lambda$ corrections can in principle enhance the $\mathcal B(u' \to t h)$ in the small $s_{tU}$ limit (in the extreme case $\mathcal B (u'\to t h)=1$ the present bound is then strengthened to $m_{u'}\gtrsim 850$~GeV~\cite{atlas-bound}) but cannot reduce it significantly below its value in the renormalizable model. 
%While a suppression of  $\mathcal B(u' \to t h)$ could still come from higher order $1/\Lambda$ corrections, they would necessarily imply a degree of fine-tuning. In the extreme case where $\mathcal B(u' \to t h ) \ll 1$ (while the $u' \to t Z$ and $u' \to b W$ rates remain as in the renormalizable model), the current direct search constraints require $m_{u'}>400$~GeV~\cite{Garberson:2013jz}. 
However, if $u'$ does not dominantly decay to third generation quarks (but instead to first two quark generations), the current direct search constraints are relaxed dramatically (c.f.~\cite{Aad:2012bt}) and $m_{u'} \simeq 300$~GeV becomes a possibility.  In summary, the Higgs fit already provides an interesting complementary constraint on  scenarios with an up-type singlet vector-like quark cancelling the top-loop quadratic divergence to the Higgs mass. Although it is at present only marginally competitive with existing direct search bounds, it is far less sensitive to the hierarchy of mixings with the known three generations of up-type quarks (provided they are small).

%%%%%%%%%%%%%%%%%%%%%%%%%%%%%%%%%%%%%%%%
%
\section{Singlet down-like vector-like quark}
\label{sec:IV}
%
%%%%%%%%%%%%%%%%%%%%%%%%%%%%%%%%%%%%%%%%

%%%%%%%%%%%%%%%%%%%%%%%%%%%%%%%%%%%%%%%%
\subsection{Renormalizable model}
\label{sec:IVa}
%%%%%%%%%%%%%%%%%%%%%%%%%%%%%%%%%%%%%%%%

Next we consider a SM extension with a vector-like quark pair $(D_L,D_R)$ in the $\bf 1_{-1/3}$ electroweak representation. The most general renormalizable Lagrangian now contains the Yukawa and mass terms 
\begin{equation}
-\mathcal{\mathcal{L}}^{(0)}_{D}= y^{ij}_{d}{\bar q^i_{L}}{H}d^j_{R} + y^{ij}_{u}{\bar q^i_{L}}\tilde{H}u^j_{R}+y^i_{D}{\bar q^i_{L}}{H}D_{R}+M_{D}{\bar D_{L}}D_{R}+\textrm{h.c.}\,.\label{eq:Lagdown}
\end{equation}
The mass matrices of up- and down-type quarks after EWSB have the form~\eqref{eq:Mu} with the replacement $u\leftrightarrow d$ and $U \leftrightarrow D$\,. In the mass-eigenbasis of 4 $(d,s,b,d')$ physical down-type and 3 $(u,c,t)$ up-type quarks the weak gauge and Higgs interactions are controlled by the general $3\times 4$ matrix $V^L_{ij}$ (again $V^R=0$) defined as before leading to $X^u = \mathbb{I}_{3\times 3}$. On the other hand, now the entries of the hermitian matrix $X^d$ are experimentally severily constrained  by their tree-level contributions to CKM non-unitarity and FCNCs in the down-quark sector and already preclude significant mixing of the vector-like down-type singlet quark with any of the SM quark generations  
\beq
|X^{d}-\mathbb I |_{3\times 3} < \left[ \begin{array}{ccc}
0.004 & 1.4\times 10^{-5} & 4 \times 10^{-4}  \\
  & 0.006 & 0.001 \\
  & & 0.0057
 \end{array} \right]\,.
\eeq
We immediately observe that Higgs phenomenology in the renormalizable down-type singlet model is again indistinguishable from the SM. In particular, considering only the dominant effects due to $b-d'$ mixing and thus parametrizing $X^d_{bb} = c_{bD}^2$, $X^d_{bd'}=c_{bD} s_{bD}$ and $X^d_{d'd'}=s_{bD}^2$, where $c_{bD}^2 + s_{bD}^2=1$, experimental constraints indicate $s_{bD}=0.05(4)$ (see Appendix~\ref{sec:Z}). This leads to maximum allowed relative deviations from SM predictions for the decay channels $h \to b\bar b$, $h \to gg$ and $h \to \gamma \gamma$ of $0.4\%$, $0.5\%$ and $-0.02\%$, respectively.

%%%%%%%%%%%%%%%%%%%%%%%%%%%%%%%%%%%%%%%%
\subsection{Non-renormalizable models}
\label{sec:IVb}
%%%%%%%%%%%%%%%%%%%%%%%%%%%%%%%%%%%%%%%%

The leading higher dimensional modifications of Higgs interactions can again be most conveniently parametrized via the replacement
\begin{align}
M_{D} &\to M_D +  c_2 \frac{v^2/2-|H|^2}{\Lambda}\,, 
\end{align}
plus an additional Higgs-dependent `kinetic mixing' operator
\beq
- \mathcal L_D^{(1)} = c^i_1 \frac{v^2/2-|H|^2}{\Lambda} \bar D_L d^i_R\,,
\eeq 
yielding new Higgs interactions in the mass eigenbasis of the form~\eqref{eq:LagHiggs1}, where now $X^{u\prime} =0$ and $X^{d\prime}  = U_L^d . [(0,0),(c_1  ,c_2 )] . U_R^{d\dagger}$\,.
Constraints on $|(U_L^d)_{4i}|^2 = 1 - X^d_{ii}$ lead to the following indirect bounds $|X^{d\prime}_{dd}| \lesssim  0.06 \,{\rm max} (c_1^i)$, $|X^{d\prime}_{ss}| \lesssim  0.08\, {\rm max} (c_1^i)$ (both allowing for sizeable modifications of $\Delta \gamma$) and $|X^{d\prime}_{bb}| \lesssim  0.13\, {\rm max} (c_1^i)$. In fact, while the $Z\to b\bar b$ anomaly cannot be fully resolved in this model, the data prefers non-zero $b-d'$ mixing with $s_{bD} = 0.05(4)$ (assuming negligible $d'$ mixing with first two generations). This is enough to allow for $\mathcal O(1)$ modification of $y_{bb}$ and thus $R_{b\bar b}$ at order $1/\Lambda$. In particular neglecting also the $m_b/M_D$ suppressed right-handed $b-d'$ mixing one can write
\beq
y_{bb} \simeq \frac{m_b}{v} + s_{bD} c_1^b \frac{v}{  \Lambda} = \frac{m_b}{v} \left(1 + s_{bD} c_1^b \frac{v^2}{m_b \Lambda}\right).
\eeq
On the other hand, $b-d'$ mixing has negligible effects on the modifications to gluon fusion production and Higgs decay to two photons
\begin{equation}
R_{gg}=\frac{0.057{}^{2}+\left|0.65+0.67y_{d'd'}\right|^{2}}{0.65^{2}},\;\; R_{\gamma\gamma}=\frac{\left|-6.5+0.45y_{d'd'}\right|^{2}}{\left|-6.5\right|^{2}},
\end{equation}
where $c_2$ contributions to $y_{d'd'}$ dominate as $y_{d'd'}=- c_{bD} c_{2}{v^{2}}/{\Lambda m_{d'}}$\,.
The resulting predictions from the non-renormalizable model with a singlet down-type
vector-like quark are presented by the red-dotted curves in the $(R_{gg},R_{\gamma\gamma})$
plane in Fig.~\ref{fig:-plot-1}. Again we have used $\left|c_{2}\right|/\Lambda\leq1\,$TeV$^{-1}$
and $m_{d'}>350\,$GeV as suggested by direct searches (see the related discussion below). 
Allowing for $R_{bb} \neq 1$ (right plot), the preferred parameter regions for $y_{d'd'}$ at $68\%$ C.L. are $y_{d'd'}=-0.16_{-0.14}^{+0.19}$ (when marginalizing over $\Delta \gamma$) and $y_{d'd'}=-0.17_{-0.13}^{+0.17}$ (when fixing $\Delta\gamma=0$). Consequently, current Higgs data are  not yet very constraining in this  context. On the other hand, in absence of significant $d'$ mixing with lighter quarks (for $R_{bb}= 1$ and $\Delta \gamma=0$ in left plot of Fig.~\ref{fig:-plot-1}), the Higgs data already give an interesting constraint (as discussed below) on  $y_{d'd'}=-0.12\pm0.08$.  Marginalizing instead over $y_{d'd'}$ and $R_{bb}$ in this scenario, we obtain a bound on $\Delta\gamma<1.2$ at $95\%$ C.L. This implies $|X_{dd}^{d\prime}+X_{ss}^{d\prime}|{v}/{\Lambda}<0.027$. Similarly, marginalizing over $y_{d'd'}$ and $\Delta\gamma$, we get $R_{bb}<2.3$, implying $|0.019-X_{bb}^{d\prime}{v}/{\Lambda}|<0.038$\,.

Finally, turning to the naturalness condition in~\eqref{eq:natural}, for the case of a single vector-like down-type singlet quark it reads in the small $b-d'$ mixing limit
\begin{equation}
\frac{m_{t}^{2} + m_{d'}^2 s_{bD}^2}{v^{2}} = c_{bD} c_2 \frac{m_{d'}}{\Lambda}+\mathcal O(1/\Lambda^{2})\,,
\end{equation}
or equivalently $y_{d'd'} = -s_{bD}^2-(m_t/m_{d'})^2$ again independent of the cut-off scale $\Lambda$. Present Higgs data then provide an indirect constraint on the $d'$ mass, which reads $m_{d'}>330$~GeV in the zero $b-d'$ mixing case and grows stronger for non-zero $s_{bD}$. This is to be compared to direct experimental searches~\cite{Garberson:2013jz}, which yield $m_{d'}> 480$~GeV for the renormalizable down-type singlet model dominantly mixing with the $b$\,. In this case however, the direct constraint is dominated by the $d' \to W t$ decay signature. Enhancing the $d' \to b h$ rate in the small $b-d'$ mixing limit through the coupling  $y_{d'b} \simeq   - c_{bD}  c'_{bD} c_1^b v/\Lambda$ can thus naturally relax it to $m_{d'}\gtrsim 350$~GeV; dominant (but small) mixing with the first two generations possibly even further. In light of this, the Higgs data already provide a complementary and competitive handle on such models.

%%%%%%%%%%%%%%%%%%%%%%%%%%%%%%%%%%%%%%%%
%
\section{Doublet vector-like quark}
\label{sec:V}
%
%%%%%%%%%%%%%%%%%%%%%%%%%%%%%%%%%%%%%%%%

%%%%%%%%%%%%%%%%%%%%%%%%%%%%%%%%%%%%%%%%
\subsection{Renormalizable model}
\label{sec:Va}
%%%%%%%%%%%%%%%%%%%%%%%%%%%%%%%%%%%%%%%%

\begin{figure}
\begin{centering}
\includegraphics[scale=1.1]{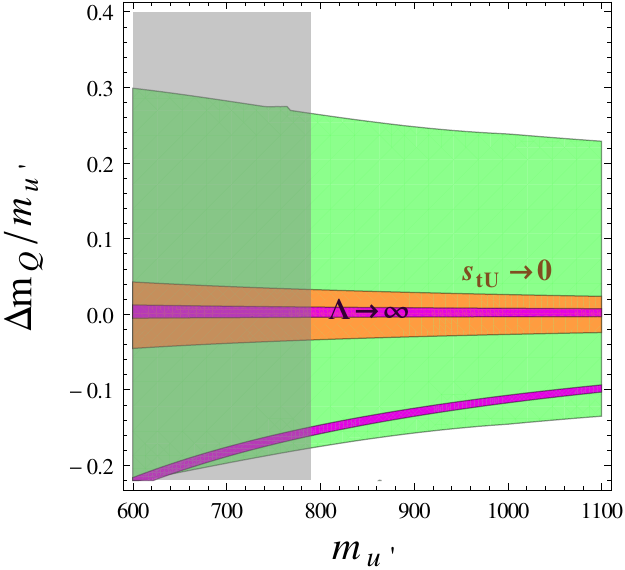}
\end{centering}
\caption{Allowed region at $95\%$ C.L. for $u' - d'$ mass splitting as a function of the $u'$ quark mass in models with a doublet vector-like quark. The narrowest purple bands apply to the renormalizable model, the middle orange band stands for the non-renormalizable model in the zero $t -u'$ mixing limit, while the broadest green band is for the non-renormalizable model with non-zero $t -u'$ mixing effects allowed by $Z\to b\overline{b}$ data. The gray area marks the ATLAS experimental search bound on the renormalizable model using the $u' \to t h$ decay signature~\cite{atlas-bound}.}
\label{fig:rhoQ2}
\end{figure}

As a final example, we consider the SM extended by a vector-like pair $(Q_L, Q_R)$ in the ${\bf 2}_{1/6}$ electroweak representation. The most general renormalizable Lagrangian now contains the Yukawa and mass terms 
\begin{equation}
-\mathcal{\mathcal{L}}^{(0)}_{Q}= y^{ij}_{d}{\bar q^i_{L}}{H}d^j_{R} + y^{ij}_{u}{\bar q^i_{L}}\tilde{H}u^j_{R}+y^i_{D}{\bar Q_{L}}{H}d^i_{R}+y^i_{U}{\bar Q_{L}}{\tilde H}u^i_{R}+M_{Q}{\bar Q_{L}}Q_{R}+\textrm{h.c.}\,.\label{eq:Lagdown}
\end{equation}
The mass matrices of both up- and down-like quarks after EWSB now have the form
\beq
\label{eq:Mdoublet}
\mathcal{M}_{u} = 
\left(\begin{array}{cc}
 y_u v/\sqrt 2\, & 0 \\
 y_U v/\sqrt 2\, & M_Q
\end{array} \right)\,,\quad
\mathcal{M}_{d} = 
\left(\begin{array}{cc}
 y_d v/\sqrt 2\, & 0 \\
 y_D v/\sqrt 2\, & M_Q
\end{array} \right)\,.
\eeq
In the quark mass eigenbasis the weak gauge and Higgs interactions of 4 ($u,c,t,u'$) physical up-like and 4 ($d,s,b,d'$) down-like quarks are governed by two $4\times 4$ matrices, a unitary  $V^L$ and a non-unitary $V^R$. Consequently $X^{u,d}_{ij} = \delta_{ij}$, while $Y^{u,d}_{ij}$ are hermitian and constrained as
\beq
|Y^u |_{3\times 3} < \left[ \begin{array}{ccc}
0.11 & 2.1 \times 10^{-4} & 0.14  \\
  & 0.018 & 0.14 \\
  & & -
 \end{array} \right]\,, \quad |Y^d |_{3\times 3} < \left[ \begin{array}{ccc}
0.1 & 1.4\times 10^{-5} & 4 \times 10^{-4}  \\
  & 0.21 & 0.001 \\
  & & 0.03
 \end{array} \right]\,.
\eeq
Due to such severe experimental bounds on the mixing of vector-like doublet components with the first two quark generations, and also with the $b$ quark, the dominant effect on Higgs phenomenology could possibly come from the mixing in the top sector (via induced $Y^u_{tt}$), which remains unconstrained at the tree-level. However, as shown in~\cite{Sally} the (right-handed) mixing angles in the top ($t-u'$) and bottom ($b-d'$) quark sectors are related via the mass splitting between the two extra quark states $u'$ and $d'$ as 
\begin{align}
\label{eq:doubletmassrelation}
m^2_{d'} [1 - s_{bD}^2 (1-r^2_{bd'})] &= m^2_{u'} [1 - s_{tU}^2 (1-r^2_{tu'})]\,,
\end{align}
where $r_{ij}\equiv m_i/m_j$\,. Left-handed and right-handed mixing angles are now related through $\tan\theta'_{ij}=r_{ij'}\tan\theta_{ij}$. At the one-loop level, the $u'-d'$ mass splitting ($\Delta m_Q \equiv m_{u'}-m_{d'}$) is constrained from EW precision measurements. 
In particular, $Z\to b\bar b$ observables constrain the $b-d'$ and $t-u'$ mixing angles as shown in Fig.~\ref{fig:tb} (see appendix~\ref{sec:Z} for details). Together with a constraint from the $\rho$ parameter, this gives the bound on $\Delta m_Q$ as shown in Fig.~\ref{fig:rhoQ2} (the narrowest purple bands).\footnote{Note that we do not find a bound as strong as reported in~\cite{Sally}. The resulting implications for Higgs phenomenology remain however qualitatively unchanged.} 
Taking all this into account (in particular discarding the fine-tuned solution for large negative $\Delta m_Q$), we find (in accordance with~\cite{Sally}) that the vector-like quark doublet with renormalizable couplings has unobservable effects in single Higgs production and decay processes.

%%%%%%%%%%%%%%%%%%%%%%%%%%%%%%%%%%%%%%%%
\subsection{Non-renormalizable models}
\label{sec:Vb}
%%%%%%%%%%%%%%%%%%%%%%%%%%%%%%%%%%%%%%%%

At the non-renormalizable level, the doublet vector-like quark model offers a somewhat richer structure than the singlet examples. Namely, the introduction of dimension five operators allows to shift the vector-like mass independently for both isospin components of $Q$ via the insertion of an iso-triplet combination of Higgs fields
\beq
M_Q \bar Q_R Q_L \to M_Q  \bar Q_R Q_L +  \frac{c^+_2}{\Lambda}  (v^2/2 -|H|^2)\bar Q_R Q_L +  \frac{c^-_2}{\Lambda} \bar Q_R  (%\sigma_3 v^2/2  + 
H H^\dagger - \tilde H \tilde H^\dagger) Q_L\,.
\eeq
Similarly, one can now introduce two new operators (via iso-singlet and iso-triplet Higgs field insertions)
\beq
-\mathcal L_Q^{(1)} =  \frac{(c^+_1)^i}{\Lambda} (v^2/2 - |H|^2) \bar Q_R q_L^i +  \frac{(c^-_1)^i}{\Lambda} \bar Q_R ( %\sigma_3 v^2/2  + 
H H^\dagger - \tilde H \tilde H^\dagger)q_L^i\,.
\eeq
The two isospin breaking corrections (proportional to $c_{1,2}^- $) now necessarily induce corrections to quark masses and mixings. In particular, the resulting changes to $\mathcal M_{u,d}$ can be parametrized as 
\begin{align}
(y_{u,d})^{ij} &\to (y'_{u,d})^{ij} \equiv (y_{u,d})^{ij} \frac{ \bar M_{U,D}}{M_{U,D}}  \mp  \frac{v^2  (y_{U,D})^i (c_1^-)^j }{2 \Lambda M_{U,D}} \,, \\
(y_{U,D})^{i} &\to  (y'_{U,D})^{i} \equiv (y_{U,D})^{i} \frac{ \bar M_{U,D}}{M_{U,D}} \pm \frac{ v^2  (y_{u,d})^{ij} (c_1^-)_j  }{ 2 \Lambda M_{U,D}} \,, \\
M_Q &\to  M_{U,D}\,,
\end{align}
where $M_{U,D} \equiv \sqrt{\bar M_{U,D}^2 + v^4 ( (c_1^-)^i (c_1^-)^*_i)/4\Lambda^2  }$ and $\bar M_{U,D} \equiv M_{Q}  \pm v^2 c_2^-/2 \Lambda$. The only observable effect of these shifts is the breaking of correlation between the masses and mixing angles in the up- and down-type quark sectors\,, such that $\Delta m_Q \equiv m_{u'}- m_{d'}$ becomes an independent free parameter, given in the zero-mixing limit (when $y'_{U,D} = 0$) solely by $\Delta m_Q =   v^2 c_2^- / \Lambda $. At the one-loop level it will affect the $\rho$ parameter as
\begin{equation}
\Delta \rho \simeq - \frac{\alpha N_C}{6 \pi s_W^2} \frac{(\Delta m_Q)^2}{m_W^2}  \,,
\end{equation}
where we have only kept the leading $\Delta m_Q$ dependence. The resulting constraint is shown in Fig.~\ref{fig:rhoQ2} (in middle orange band). However, if we also include non-zero $t-u'$ mixing effects, marginalizing over the allowed values of $s_{tU}$ from $Z\to b\bar b$ data we obtain a much weaker bound on $\Delta m_Q$ shown in the uppermost (green) band in Fig.~\ref{fig:rhoQ2}.  In our numerical evaluation we employ the full one-loop formula for $\Delta\rho$, which can be found in Appendix~\ref{sec:rhoDoublet}.
We thus conclude that  after including the contributions of leading higher dimensional operators, the isospin components of a TeV scale vector-like quark doublet can be split by as much as $30\%$. Although this has no observable consequences for Higgs phenomenology, it can have profound implications for direct $u'$ searches if the $u' \to d' W$ decay channel becomes kinematically allowed.

The new Higgs interactions in the mass eigenbasis are again of the form~\eqref{eq:LagHiggs1}, where now $X^{u,d\prime} = U_R^{u,d} . [(0,0),(c^+_1 \pm c^-_1  ,c_2^+ \pm c^-_2 )]^* . U_L^{u,d\dagger}$ or explicitly $(X^{u,d\prime})_{ij} = (U_R^{u,d})_{i4} [(c^+_1 \pm c^-_1)^k (U_L^{u,d})^*_{jk} + (c_2^+ \pm c^-_2)  (U_L^{u,d})^*_{j4}  ]$. They are thus  constrained indirectly by bounds on $Y^{u,d}$, since now $|(U_R^{u,d})_{i4}|^2 = Y^{u,d}_{ii}$. Taking into account also the relation $|(U_L^{u,d})_{i4}|^2 = (m_i/M_{U,D})^2 | (U_L^{u,d})_{i4} |^2$, we can safely neglect $c_2^\pm$ contributions and obtain $X^{u\prime}_{uu} \lesssim 0.35\,{\rm max}[(c^+_1 + c^-_1)^i]$, $X^{u\prime}_{cc} \lesssim 0.13\,{\rm max}[(c^+_1 + c^-_1)^i]$, $X^{d\prime}_{dd} \lesssim 0.30\,{\rm max}[(c^+_1 - c^-_1)^i]$, $X^{d\prime}_{ss} \lesssim 0.40\,{\rm max}[(c^+_1 - c^-_1)^i]$ and $X^{d\prime}_{bb} \lesssim 0.11\,{\rm max}[(c^+_1 - c^-_1)^i]$.  Significant effects in $\Delta \gamma$ and $R_{bb}$ are thus possible and can be used to constrain the diagonal entries of $X^{u, d\prime}$\,. 

Additional interesting effects again appear in loop induced Higgs processes. Higgs decays to pairs of gluons or photons are modified by additional heavy particles in the loop
\begin{align}
R_{gg}&=\frac{\left|0.68 (r_x+r_y)-0.040\right|^{2}+\left.0.057\right.^{2}}{0.65^{2}}\,,\quad
R_{\gamma\gamma}=\frac{\left|-8.3+1.8r_x+0.45r_y\right|^{2}}{\left|-6.5\right|^{2}}\,,
\end{align}
where to order $1/\Lambda$ and including $t-u'$ mixing
  \begin{align}
 r_{x} & \equiv y_{tt}\frac{v}{m_{t}}+y_{u'u'}\frac{v}{m_{u'}} \notag \\
  & =1+s_{tU}\left((c_{1}^{+}+c_{1}^{-})^{t}c'_{tU}-(c_{2}^{+}+c_{2}^{-})s'_{tU}\right)\frac{v^{2}}{\Lambda m_{t}}-c_{tU}\left((c_{1}^{+}+c_{1}^{-})^{t}s'_{tU}+(c_{2}^{+}+c_{2}^{-})c'_{tU}\right)\frac{v^{2}}{\Lambda m_{u'}}\,,
\end{align}
and
\begin{align}
 r_{y} \equiv y_{d'd'}\frac{v}{m_{d'}}=-\frac{v^{2}}{\Lambda m_{d'}}(c_{2}^{+}-c_{2}^{-})\,.
\end{align}
Taking into account the bound on $t-u'$ mixing, a good approximation for $r_{x}=1+s_{tU}\frac{(c_{1}^{+}+c_{1}^{-})^{t}v^{2}}{\Lambda m_{t}}-\frac{(c_{2}^{+}+c_{2}^{-})v^{2}}{\Lambda m_{u'}}$. There is no correlation between $R_{gg}$ and $R_{\gamma \gamma}$ in general, unless one imposes additional constraints on the parameters. 

Therefore, the resulting predictions from the non-renormalizable model with a doublet vector-like quark are given by a region (dashed-black contour) in the $(R_{gg},R_{\gamma \gamma})$ plane of Fig.~\ref{fig:-plot-1} left, in the case when modification of $\Delta\gamma$ is allowed and $R_{bb}=1$, and on Fig.~\ref{fig:-plot-1} right, when sizable modification of $R_{bb}$ is allowed as well.\footnote{As in the case of the non-renormalizable model with a down-like vector-like quark, $y_{bb}$ can receive $\mathcal O(1)$ modifications even for small $b - d'$ mixing.} We have assumed $\left|c_{1,2}^{+,-}\right|/\Lambda\leq1\,\textrm{TeV}^{-1}$ and $m_{q'}>790\,\textrm{GeV}$
as suggested by direct searches~\cite{atlas-bound}. Also, we take the $t-u'$ mixing angle to be within the $95\%$ C.L. experimental bound discussed above ($s_{tU}\lesssim 0.35$).

For small mixing and negligible isospin breaking, masses of $u'$ and $d'$ quarks are degenerate and $r_y=r_x-1=-\frac{v^{2}}{\Lambda m_{q'}}c_{2}^{+}$.   Allowing for modification
of $\Delta\gamma$ and fixing $R_{bb}=1$, the resulting predictions in this scenario are presented by the green-dashed
curve in the $(R_{gg},R_{\gamma\gamma})$ plane in Fig.~\ref{fig:-plot-1} left. Allowing for sizable modification of $R_{bb}$ as well, the predictions are presented by the green-dashed curve in the $(R_{gg},R_{\gamma\gamma})$ plane in Fig.~\ref{fig:-plot-1} right. Again we have assumed $\left|c_{2}^{+}\right|/\Lambda\leq1\,\textrm{TeV}^{-1}$ and $m_{q'}>790\,\textrm{GeV}$.  In particular,
the preferred parameter regions for $r_{y}$ at $68\%$ C.L. are  $r_{y}=-0.09_{-0.06}^{+0.09}$ (marginalizing over both $R_{bb}$ and $\Delta \gamma$), $r_{y}=-0.09_{-0.06}^{+0.08}$ (marginalising over $R_{bb}$ but fixing $\Delta \gamma$), $r_{y}=-0.07_{-0.04}^{+0.09}$ (fixing $R_{bb}=1$ and marginalizing over $\Delta \gamma$) and finally $r_{y}=-0.06\pm 0.04$ (fixing both $R_{bb}$ and $\Delta \gamma$ to their SM values) . 

Also in the unmixed isospin symmetric doublet scenario we can get $\Delta\gamma<1.0$ at $95\%$ C.L. after marginalizing over $r_{y}$ and $R_{bb}$. This implies $|X_{uu}^{u\prime}+X_{cc}^{u\prime}+X_{dd}^{d\prime}+X_{ss}^{d\prime}|{v}/{\Lambda}<0.025$. Conversely, marginalizing over $r_{y}$ and $\Delta\gamma$, we get $R_{bb}<2.1$ implying $|0.019-X_{bb}^{d\prime}{v}/{\Lambda}|<0.036$\,. We have checked that these results remain stable even in presence of small $t-u'$ mixing and isospin breaking.

In this scenario, the Higgs mass naturalness condition reads
\begin{equation}
\frac{m_{t}^{2}}{v^{2}}=2c_{2}^+\frac{m_{q'}}{\Lambda}+\mathcal{O}(1/\Lambda^{2})\,,
\end{equation}
or equivalently $r_y=-(m_{t}/\sqrt{2}m_{q'})^{2}$. This condition allows to put an indirect bound on the mass of $q'$ to be $390\,$GeV at $95\%$ C.L. in the case of fixed $R_{bb}$.

Turning to direct searches, the most severe bound on $u'$ in the renormalizable model with a doublet vector-like quark assuming dominant but small mixing with the third generation, $m_{u'}>790$~GeV~\cite{atlas-bound} is given by the ATLAS experimental search using the $u' \to t h$ decay signature. In the non-renormalizable model the relevant couplings are given by $y_{tu'} =  s_{tU} c_{tU} m_{u'}/v +s_{tU}  (s'_{tU} (c_1^++c_1^-)^t + c'_{tU}(c_2^++c_2^-) )v/\Lambda$ and $y_{u't} =  s_{tU} c_{tU} m_{t}/v - c_{tU}  (c'_{tU}  (c_1^++c_1^-)^t -  s'_{tU}(c_2^++c_2^-) )v/\Lambda$. It is then easy to check that again compared to $u' \to t Z$ and $u' \to b W$ rates, the $1/\Lambda$ corrections can in principle enhance $\mathcal B(u' \to t h)$ in the small $s_{tU}$ limit but cannot reduce it significantly below its value in the renormalizable model. However, the significant $u'-d'$ mass splitting allowed by present data when including dimension five contributions, reopens the possibility that the dominant decay channel of $u'$ is actually $u' \to d' W$, in which case the existing direct search constraints are considerably relaxed and dominated by searches for the lighter isospin component via $d' \to t W$ and $d' \to b h$ decay signatures (in the case of dominant mixing with the third generation). Consequently, in such scenarios $u' (d')$ masses as low as $m_{u'(d')} \gtrsim 400$~GeV\,($300$~GeV) could still be viable.

%%%%%%%%%%%%%%%%%%%%%%%%%%%%%%%%%%%%%%%%
%
\section{Conclusions}
\label{sec:VI}
%
%%%%%%%%%%%%%%%%%%%%%%%%%%%%%%%%%%%%%%%%

We have  systematically investigated the  impact of  dynamical vector-like quarks, accommodated within SM gauge representations and charges, on Higgs physics. In particular, we have considered the weak singlet up-type, singlet down-type and doublet vector-like quarks, potentially mixing with all three known generations of chiral quarks, and have updated the most relevant constraints on such scenarios from low energy flavour phenomenology and electro-weak precision measurements. The resulting general constraints on the renormalizable and leading non-renormalizable quark-Higgs interactions are summarised in Tables~\ref{tab1} and~\ref{tab2}, respectively.

\begin{table}
\begin{centering}
\begin{tabular}{|c|l|c|}
\hline
Coupling & Constraint & Reference \tabularnewline
\hline
\hline
$|X^u_{cu}|, |Y^u_{cu}|$ & $ < 2.1\times 10^{-4}$ & \cite{Golowich:2007ka,hfag} \tabularnewline
\hline
$|X^u_{tu,tc}|,|Y^u_{tu,tc}|$ & $ < 0.14$ & Appendix~\ref{sec:top} \tabularnewline
\hline
${\rm }|X^d_{ds}|, {\rm }|Y^d_{ds}|$ & $ < 1.4\times 10^{-5}$ & {\multirow{3}{*}{Appendix~\ref{sec:down}} } \tabularnewline
\cline{1-2}
$|X^d_{db}|,|Y^d_{db}|$ & $ < 4 \times 10^{-4}$ &  \tabularnewline
\cline{1-2}
$|X^d_{sb}|,|Y^d_{sb}|$ & $ < 1\times 10^{-3}$ &  \tabularnewline
\hline
\hline
$\delta X_{uu}^{u}$ & $-0.0001(6)$ & \cite{Dowdall:2013rya} \tabularnewline
\hline
$\delta X_{cc}^{u}$ & $-0.0020(13)$ & Appendix~\ref{sec:Z} \tabularnewline
\hline
\hline
$\delta  X_{dd}^{d}$ & $-0.0031(20)$ & {\multirow{3}{*}{Appendix~\ref{sec:Z}} } \tabularnewline
\cline{1-2}
$\delta  X_{ss}^{d}$ & $-0.002(3)$ & \tabularnewline
\cline{1-2}
$\delta  X_{bb}^{d}$ & $0.0027(15)$ & \tabularnewline
\hline
\hline
$\delta Y_{uu}^{u}$ & $0.035(40)$ & {\multirow{5}{*}{Appendix~\ref{sec:Z}} } \tabularnewline
\cline{1-2}
$\delta Y_{cc}^{u}$ & $-0.003(9)$ & \tabularnewline
\cline{1-2}
$\delta Y_{dd}^{d}$ & $0.030(35)$ & \tabularnewline
\cline{1-2}
$\delta Y_{ss}^{d}$ & $-0.05_{-0.06}^{+0.08}$ & \tabularnewline
\cline{1-2}
$\delta Y_{bb}^{d}$ & $-0.018(6)$ & \tabularnewline
\hline
\end{tabular}
\par\end{centering}
\caption{\label{tab1}Compilation of phenomenological constraints on Higgs (and $Z$) couplings to SM quarks in renormalizable vector-like quark models (see Eqs.~\eqref{eq:LZ},\,\eqref{eq:Lh0}) from precision flavor and electroweak observables. All upper bounds are given at 95\% C.L. For discussion of $\delta X^u_{tt}$, $\delta Y^u_{tt}$ constraints see Sections~\ref{sec:IIIa} and~\ref{sec:Va}, respectively.}
\end{table}

\begin{table}
\begin{centering}
\begin{tabular}{|c|l|c|}
\hline
Coupling & Constraint & Reference\tabularnewline
\hline
\hline
$|X^{u\prime} _{uc,cu}|v/\Lambda$ & $<7\times10^{-5}$ & {\multirow{5}{*}{\cite{Harnik:2012pb}} } \tabularnewline
\cline{1-2}
$\sqrt{|X^{u\prime} _{tu,tc}|^2 + |X^{u\prime} _{ut,ct}|^2}v/\Lambda$ & $<0.34$ &  \tabularnewline
\cline{1-2}
$|X^{d\prime}_{sd,ds}|v/\Lambda$ & $<2\times10^{-5}$ &  \tabularnewline
\cline{1-2}
$|X^{d\prime}_{bd,db}|v/\Lambda$ & $<2\times10^{-4}$ &  \tabularnewline
\cline{1-2}
$|X^{d\prime}_{sb,bs}|v/\Lambda$ & $<1\times10^{-3}$ &  \tabularnewline
\hline
\hline
$\sqrt{|X_{uu}^{u'}|^2+|X_{cc}^{u'}|^2}{v}/{\Lambda}$ & $<0.022$ & {\multirow{1}{*}{Section~\ref{sec:IIIb}} } \tabularnewline
\hline
$\sqrt{|X_{dd}^{d'}|^2+|X_{ss}^{d'}|^2}{v}/{\Lambda}$ & $<  0.027$ & {\multirow{1}{*}{Section~\ref{sec:IVb}} } \tabularnewline
\hline
$\sqrt{|X_{uu}^{u'}|^2+|X_{cc}^{u'}|^2+|X_{dd}^{d'}|^2+|X_{ss}^{d'}|^2}{v}/{\Lambda}$ & $ < 0.025$ &  {\multirow{1}{*}{Section~\ref{sec:Vb}} } \tabularnewline
\hline
\multirow{2}{*}{$|0.019-X_{bb}^{d'}{v}/{\Lambda}|$} &  {$<  0.038$} &  {Section~\ref{sec:IVb}} \tabularnewline
 &  {$<  0.036$} &  {Section~\ref{sec:Vb}} \tabularnewline
\hline
\end{tabular}
\par\end{centering}
\caption{\label{tab2} Compilation of constraints on additional Higgs couplings to SM quarks due to dimension five operators in models with vector-like quarks (see Eq.~\eqref{eq:LagHiggs1}). The off-diagonal couplings are bounded by low energy flavor observables, whereas upper limits on the diagonal ones were estimated from the fit to current Higgs data. All  bounds are given at 95\% C.L. For discussion of $\delta X^{u\prime}_{tt}$ constraints within up-type singlet and doublet vector-like quark scenarios see Sections~\ref{sec:IIIb} and~\ref{sec:Vb}, respectively.}
\end{table}

Within the renormalizable SM extended by additional vector-like quarks, we have shown generally that Higgs couplings to the known three generations of quarks need to remain SM-like regardless of the extra quark masses.  This feature is a consequence of the fact that precision flavour and electro-weak observables are affected by the mixing of vector-like and chiral quarks, some of them already at the tree level. 
Consequently, Higgs decay widths to light quark pairs, $ gg$ and $\gamma \gamma$ cannot  deviate significantly from their SM predictions.

A singular feature of models with vector-like fermions is that non-renormalizable contributions sensitive to physics at the cut-off scale of the effective low energy theory appear already at dimension five. The inclusion of such higher-dimensional operators is essential in models that aim to cancel dominant quadratic divergences to the Higgs boson mass coming from top quark loops with new fermionic contributions. 
Contrary to the renormalizable models (see however~\cite{Bonne:2012im}), they also predict interesting effects in Higgs phenomenology. We have investigated such contributions for all three types of  additional  vector-like quarks mixing with SM generations (see Fig.~\ref{fig:-plot-1}). The most important consequences for Higgs physics are possible significant enhancements in Higgs decay rates to pairs of light ($u\bar u, d\bar d, s\bar s, c\bar c$) quarks, which may still account for a significant fraction of the total Higgs decay width. This is possible due to a simultaneous strong modification of the Higgs production by gluon fusion $gg \to h$ (up to $50\%$ deviation from SM predictions,
while the $h\to \gamma \gamma$ decay width can receive modifications up to $10\%$). Such a possibility could thus be tested by future more accurate determinations of vector boson fusion and ($W$, $Z$ or $t\bar t$) associated Higgs production.In addition, a possible modification of the Higgs coupling to $b$ quarks in models containing down-type vector-like quarks will be probed by future precise measurements of the $h\to b\bar b$ decay.\footnote{A direct determination of Higgs decays to charmed and light jets would  also be enlightening in this respect.}

Interestingly, current Higgs data are perfectly consistent with (and even exhibit a slight preference for) the possibility that vector-like quarks contribute to the cancellation of the top loop quadratic divergence to the Higgs mass. Conversely in some cases, a fit to existing Higgs measurements under such an assumption already offers competitive and robust constraints on vector-like quark masses in comparison with results of direct experimental searches which need to rely on particular decay signatures. In the example of a weak doublet of vector-like quarks, we have shown that dimension five contributions allow to relax the stringent bounds on the mass splitting between the two isospin states. Consequently the decay of the heavier doublet component to the lighter one with the emission of a $W$ boson may become kinematically allowed, affecting the relevant experimental signatures. More generally in presence of dimension five contributions, vector-like quark decay widths are naturally dominated by decay channels involving the Higgs.  Future dedicated experimental searches for such particular signatures (as exemplified in~\cite{atlas-bound}, see also~\cite{Azatov:2012rj}) could thus shed light on the relevance of vector-like quarks in the solution to the SM hierarchy problem.

\begin{acknowledgments}
This work was supported in part by the Slovenian research agency and the Ad futura Programme of the Slovenian Human Resources and Scholarship Fund.
\end{acknowledgments}

\appendix

%%%%%%%%%%%%%%%%%%%%%%%%%%%%%%%%%%%%%%%%
%%%%%%%%%%%%%%%%%%%%%%%%%%%%%%%%%%%%%%%%
%%%%%%%%%%%%%%%%%%%%%%%%%%%%%%%%%%%%%%%%

%%%%%%%%%%%%%%%%%%%%%%%%%%%%%%%%%%%%%%%%
%
\section{CKM non-unitarity and $Z$ mediated FCNCs in top quark production and decays}
\label{sec:top}
%
%%%%%%%%%%%%%%%%%%%%%%%%%%%%%%%%%%%%%%%%

In absence of right-handed charged currents, experimental constraints on $\sum_{i=d,s,b} |V^L_{ti}|^2= 1- \Delta^u_t \leq 1$ and also $|V^L_{tb}|^2$ can be obtained from the measurements of the ($t$-channel) single top production cross-section ($\sigma_t$) at the LHC and the fraction of top decays to $W b$ pairs in $t\bar t$ production ($R$). Assuming $t\to W q$ channels dominate the top decay width, to a very good approximation $R$ is given by $R \simeq |V^L_{tb}|^2/(\sum_{i=d,s,b} |V^L_{ti}|^2)$ and we can use the recent CMS result $R^{\textrm{exp}} = 0.98 \pm 0.04$~\cite{CMS-PAS-TOP-12-035}. Note that this measurement alone requires that $|V^L_{tb}|\gg |V^L_{ts}|, |V^L_{td}|$. The relevant $t$-channel single top production cross-section can then be written as $\sigma_t \simeq \sigma_t^{\textrm{SM}} |V^L_{tb}|^2 R$, where the SM prediction of $\sigma_t^{\textrm{SM}}= 64.6^{+2.7}_{-2.0}$~pb~\cite{Kidonakis:2011wy} is obtained with $|V^L_{tb}|=1$ and $R=1$. We compare this to the weighted average of the recent ATLAS~\cite{Aad:2012ux}, and CMS ~\cite{Chatrchyan:2012ep} results ($\sigma_t^{\textrm{exp}}=68.5 \pm 5.8$~pb\,). Performing a $\chi^2$ fit of the two experimental quantities in terms of $|V^L_{tb}|^2$ and $\sum_{i=d,s,b} |V^L_{ti}|^2$ (and taking into account theoretical constraints $\sum_{i=d,s,b} |V^L_{ti}|^2\leq 1$ and $|V^L_{ti}|>0$) we obtain the $95$\%~C.L. lower bounds of
\begin{align}
|V^L_{tb}|^2 &> 0.85\,,  \quad
\sum_{i=d,s,b} |V^L_{ti}|^2 > 0.87\,. \label{vtiB}
\end{align}

In models with up-type weak singlet vector-like quarks, $|X_{ut}|$ and $|X_{ct}|$ will contribute to FCNC top decays. Combined with $\sigma_t$ and $R$, the experimental bounds on $\mathcal B(t\to Z q)$ can then be used to constrain $\sqrt{|X_{ut}|^2+|X_{ct}|^2} \equiv |X_{tu,tc}|$. The presence of these new decay channels in principle also needs to be accommodated in $\sigma_t$ and $R$ by writing 
\begin{equation}
\label{singlet}
\sigma_t^{\textrm{}} = \sigma_{\textrm{SM}} |V^L_{tb}|^2 \left({\frac{1}{R} + \rho_{WZ} \frac{ |X_{tu,tc}|^2}{|V^L_{tb}|^2}}\right)^{-1}\,,
\end{equation}
where
\begin{equation}
\rho_{WZ} = \frac{1}{2} \frac{(2 M_Z^2 + m_t^2) \left(1-\frac{M_Z^2}{m_t^2}\right)^2}{(2 M_W^2 + m_t^2) \left(1-\frac{M_W^2}{m_t^2}\right)^2},
\end{equation}
takes into account the dominant phase-space difference in $t\to Wq$ and $t\to Zq$ decays. In addition, searches for $t\to Zq$ typically assume $\mathcal{B}(t \to Wb) + \mathcal{B}(t \to Zq) =1$ and $|V^L_{tb}|=1$. In presence of  $|V^L_{tb}|<1$, the experimental results should instead be compared to
\begin{equation}
\overline{\mathcal{B}}(t \to Zq) = \left({1+ \frac{|V^L_{tb}|^2}{ \rho |X_{tu,tc}|^2}}\right)^{-1}\,.
\end{equation}
Including the recent ATLAS result $\mathcal{B}(t \to Zq)^{\textrm{exp}}<0.73\%$ \cite{Aad:2012ij} in our fit, we first observe that the presence of FCNCs has no observable effect on the results in eq.~\eqref{singlet}. On the other hand, we can obtain an upper bound on $|X_{tu,tc}|<0.14$ at $95\%$~C.L.\,. 

Although the presence of right-handed charged currents complicates the analysis of top production and decays, a robust bound on $|Y_{tu,tc}|$ and also $|V^R_{tb}|$ can nonetheless be obtained. To a good approximation namely in this case
\beq
R = \frac{|V^L_{tb}|^2+|V^R_{tb}|^2}{\sum_{i=d,s,b} (|V_{ti}^L|^2+|V_{ti}^R|^2)}\,,
\eeq
and thus experimentally $|V^L_{tb}|^2+ |V^R_{tb}|^2 \gg |V^L_{ts}|^2,|V^R_{ts}|^2, |V^L_{td}|^2,|V^R_{td}|^2$\,. In addition, the presence of $V^R_{tb}$ will affect single top production as
\begin{equation}
\sigma_t^{\textrm{}} = \sigma_{\textrm{SM}} (|V^L_{tb}|^2 + \kappa_R |V^R_{tb}|) \left({\frac{1}{R} + \rho_{WZ} \frac{ |Y_{tu,tc}|^2}{|V^L_{tb}|^2+|V^R_{tb}|^2}}\right)^{-1}\,,
\end{equation}
where $\kappa_R\simeq 0.92$~\cite{AguilarSaavedra:2008gt}\,. Finally, $V^R_{tb}$ contributes to the positive $W$ helicity fraction ($\mathcal F_+$) in top decays as
\beq
\mathcal F_+ =  \frac{|V^L_{tb}|^2}{|V^L_{tb}|^2+|V^R_{tb}|^2} \mathcal F_+^{\rm SM} +  \eta^R_+ \frac{|V^R_{tb}|^2}{|V^L_{tb}|^2+|V^R_{tb}|^2} \frac{2 x}{(1+2 x)}\,,
\eeq
where $\mathcal F_+^{\rm SM} = 0.0017(1)$~\cite{Fischer:2000kx}, $x=(m_W/m_t)^2$ and $\eta^R_+ = 0.93$ parametrizes NLO QCD corrections~\cite{Drobnak:2010ej}. Using the recent determination of $\mathcal F_+^{\rm exp} = 0.01(5)$ by ATLAS~\cite{Aad:2012ky} we obtain $ |V^R_{tb}| < 0.2 |V^L_{tb}|$. Thus we may conservatively use the results of the previous paragraphs also to constrain $|V^L_{tb}|>0.85$ and $|Y_{tu,tc}|<0.14$\,.

%%%%%%%%%%%%%%%%%%%%%%%%%%%%%%%%%%%%%%%%
%
\section{Bounds on down-type quark mixing with vector-like weak singlets and doublets from rare $K$ and $B_q$ processes}
\label{sec:down}
%
%%%%%%%%%%%%%%%%%%%%%%%%%%%%%%%%%%%%%%%%

Model independent bounds on the off-diagonal entries of $X^{d}, Y^{d}$ in models with additional vector-like down-type weak singlet and weak doublet quarks respectively can be obtained from their tree-level ($Z$-mediated) contributions to FCNCs involving down-type quarks. For example, a bound on $X^d_{ds}, Y^d_{ds}$ can be extracted from the $K_L \to \mu^+ \mu^-$ decay. We use a conservative estimate for the pure short distance branching fraction $\mathcal B(K_L \to \mu^+ \mu^-)_{SD}^{\rm exp} < 2.5 \times 10^{-9}$, obtained using
dispersive techniques~\cite{Isidori:2003ts}, as a $1\sigma$ upper bound. Neglecting the much smaller SM contributions, the $X^d_{ds}$  contribution can be written as
\beq
 \mathcal B(K_L \to \mu^+ \mu^-)^{X} = \frac{G_F^2}{16\pi} f_K^2 m_K \tau_{K_L} m_\mu^2 \sqrt{1-\frac{4m_\mu^2}{m_K^2}} {\rm Re}(X_{sd})^2\,.
\eeq
Using the inputs for $G_F$, masses and lifetimes from~\cite{pdg} and also $f_K = 155.37(34)$~MeV~\cite{Dowdall:2013rya} we obtain ${\rm Re}(X_{sd})< 1\times  10^{-5}$ (the same result applies also to $Y_{sd}$)\,. Since much stricter bounds are expected on ${\rm Im}(X_{sd})$ (${\rm Im}(Y_{sd})$) from the precise knowledge of $\epsilon_K$, we interpret the above values as conservative constraints also on the moduli of $X_{sd}$ and $Y_{sd}$.

In the $B_d$ sector, $X_{bd}$ contributes at the tree-level both to $B_d \to \mu^+ \mu^-$, as well as in $B^0-\bar{B^0}$ mixing. Neglecting $X_{bd}$ loop level corrections due to CKM non-unitarity and dynamical vector-like quarks running in the loop, the $B_d \to \mu^+ \mu^-$ branching fraction can be written as
\begin{align}
\label{eq:Bdmumu}
 \mathcal B(B_d \to \mu^+ \mu^-) &= \frac{G_F^2}{8\pi} f_{B_d}^2 m_{B_d} \tau_{B_d} m_\mu^2 \sqrt{1-\frac{4m_\mu^2}{m_{B_d}^2}} \left| \lambda^t_{bd} C_{dB=1}^{\rm SM} + \frac{X_{bd}}{\sqrt 2} \right|^2\,, 
\end{align}
with $\lambda^t_{bd} = V^L_{td} V^{L*}_{tb}$ the relevant CKM combination and the SM Wilson coefficient given by
\beq
C_{dB=1}^{\rm SM}=  \frac{\alpha_{}}{\sqrt 2 \pi s_W^2} \eta_Y Y_0\left(\frac{\overline{m}_t^2}{m_W^2}\right)\,.
\eeq
Here $Y_0$ is the relevant Inami-Lim loop function~\cite{Inami:1980fz} 
\beq
Y_0(x) =\frac{x}{8} \left[ \frac{4-x}{1-x} + \frac{3x}{(1-x)^2} \ln x  \right]\,,
\eeq
while $\eta_Y = 1.01$~\cite{Buras:2012ru} parametrizes higher order QCD corrections. For the values of SM input parameters (in particular $\alpha_{\rm em}$, $s_W$ and $\overline m_t$) we follow the prescription of~\cite{Buras:2012ru}\,, and we use $f_B=190.6\pm4.7$~\cite{Laiho:2009eu}\,.
In the $B^0-\bar{B^0}$ system, the two most relevant observables are the mass-difference between the two $B_d$ mass eigenstates ($\Delta m_d$) and the CP violating phase in the mixing ($\beta_d$). Since the corresponding width difference ($\Delta \Gamma_d$) is small $ |\Delta \Gamma_d| \ll |\Delta m_d|$, one can write
\begin{align}
\label{eq:dB=2}
\Delta m_d & \simeq 2 |M^d_{12}|\,, \quad  \sin2\beta_d = \frac{\textrm{Im}(M^d_{12})}{|M^d_{12}|}  \,, \quad {\rm where} \quad M^{d}_{12} = \frac{G_F}{12}  m_{B_d} f_{B_d}^2 {B}_{B_d}\  \left( {\lambda^t_{bd}}^2 C_{dB=2}^{\textrm{SM}} + \frac{X_{bd}^2}{\sqrt{2} } \right)\,.
\end{align}
with the SM Wilson coefficient given by
\begin{equation}
C^{\textrm{SM}}_{dB=2} = \frac{\alpha_{}}{\sqrt 2 \pi s_W^2}  {\eta}_B S\left(\frac{\overline{m}_t^2}{m_W^2}\right).
\end{equation}
Here again $S$ is the relevant Inami-Lim loop function~\cite{Inami:1980fz} 
\beq
S(x) = \frac{x}{2} \left[ \frac{1}{2} + \frac{3}{2}
   \frac{1-3x}{(1-x)^2} -    \frac{3 x^2}{(1-x)^3} \ln x \right]
     \,,
\eeq
while $\eta_B = 0.939$~\cite{Lenz:2010gu} parametrizes higher order QCD corrections. In addition we use $f_{B_d}^2 {B}_{B_d}=0.0411 \pm 0.0075$~\cite{Bouchard:2011xj}.

In absence of CKM unitarity, we need to determine not only $X_{bd}$ but also the CKM combination $\lambda^t_{bd}$. For this purpose we use the radiative rate $B\to X_d \gamma$, which is unaffected by $X_{bd}$ at the tree-level. Although it receives non-standard contributions proportional to $X_{bd}$ due to CKM non-unitarity and extra down-type quarks in the loops, these are parametrically (loop) suppressed compared to tree-level effects in $B_d \to \mu^+ \mu^-$ and $B^0-\bar{B^0}$ mixing. Using the SM recent evaluation~\cite{Crivellin:2011ba}
\beq
\mathcal B(B\to X_d \gamma)^{\rm SM}_{E_\gamma>1.6\rm GeV} =  \left| \frac{\lambda^t_{bd}}{0.0084} \right|^2 1.54(26) \times 10^{-5}\,,
\eeq
and comparing it to the experimental result $\mathcal B(B\to X_d \gamma)^{\rm exp}_{E_\gamma>1.6\rm GeV} = 1.41(57)\times 10^{-5} $~\cite{delAmoSanchez:2010ae} we extract $|\lambda^t_{bd}|= 8.0_{-2.0}^{+1.6} \times 10^{-3}$\,. Plugging this into eq.~\eqref{eq:Bdmumu} we observe that compared to the experimental 95\% C.L. upper limit of $\mathcal B(B_d \to\mu^+\mu^-)^{\rm exp} <9.4 \times 10^{-10}$~\cite{Aaij:2012nna} the $\lambda^t_{bd}$  contribution can be safely neglected and we obtain a bound on $|X_{bd}| < 4\times 10^{-4} $. Note that the measurements of $\Delta m_d^{\rm exp} = 0.507(4)$~ps${}^{-1}$ and $\sin2\beta^{\rm exp}=0.679(20)$~\cite{hfag} cannot be used to impose a stricter constraint since the phases of $\lambda_{bd}^t$ and $X_{bd}$ can always be arranged such that cancellations between these contributions weaken the prospective bounds above the one by $\mathcal B(B_d \to\mu^+\mu^-)^{\rm exp}$\,.

Finally in the $B_s$ sector we can proceed similarly, with obvious replacements $\lambda^t_{bd} \to \lambda^t_{bs}$ and $B_{d} \to B_s$. We again employ $B\to X_s \gamma$ to extract $|\lambda^t_{bs}|$ (modulo loop-suppressed $X_{bs}$ effects) as $|\lambda^t_{bs}| =0.043 (2)$~\cite{pdg} and use $f_{B_s} = 227.6\pm 5.0 $~\cite{Laiho:2009eu}, $f_{B_s}^2 {B}_{B_s} = 0.0559(68)$~\cite{Bouchard:2011xj} for hadronic inputs. By comparing to the experimental values of $\Delta m_s^{\rm exp} = 17.719(43)$~ps${}^{-1}$, $ 2\beta_s^{\rm exp}= -0.1(6.1)^{\circ}$\,~\cite{hfag} and $\mathcal B(B_s \to \mu^+ \mu^-)^{\rm exp} = (3.2^{+1.5}_{-1.2})\times 10^{-9}$~\cite{Aaij:2012nna} we thus obtain a bound on $|X_{bs}|<1\times 10^{-3}$\,, which is again dominated by the muonic $B_s$ decay rate.

The presence of vector-like weak doublet quarks induces right handed charged and neutral currents among SM quarks. The resulting flavor phenomenology is very rich and deserves a dedicated study.  For our purpose however, it suffices to show that to a first approximation, one can actually neglect all terms coming from right handed charged current operators as well as the ones containing extra $u',d'$ quarks in the loops contributing to quark FCNCs. This requires some knowledge of the mixing matrices, which can be expressed through the parameters of the Yukawa sector in terms of an expansion in ratios of mass parameters, which enables us to connect the left and right handed mixing matrices. However, the approximation only holds if such an expansion is justified, as we will check a posteriori. First recall that the up- and down-like quark mass matrices in presence of a single vector-like quark doublet can be written in the form~\eqref{eq:Mdoublet}.
Diagonalizing the products $\mathcal{M} \mathcal{M}^{\dagger}$ and $\mathcal{M}^{\dagger} \mathcal{M}$ one obtains the left and right mixing matrices, respectively, which we write in the form (c.f.~\cite{Branco:1999fs})
\begin{equation}
 U^{u,d}_{L,R} = \left( \begin{array}{cc}
  K^{u,d}_{L,R} & R^{u,d}_{L,R} \\
  S^{u,d}_{L,R} & T^{u,d}_{L,R}
 \end{array} \right). \label{rot}
\end{equation}
The mixing matrices between the vector-like and chiral quarks can most easily be obtained by starting from a basis of right handed chiral quarks, where $y_{u,d} y_{u,d}^\dagger v^2/2$ are both diagonal, which can be done as the right handed chiral quarks are isosinglets. The mixing $y_{U,D} v/\sqrt 2$ is bounded by the EW scale $v$, whereas the Dirac mass $m_Q$ is experimentally required to be larger. Taking thus $v /M_Q$ as the expansion parameter,   it can be shown that, to first order in this expansion, both $4 \times 4$ right-handed rotation matrices schematically take the form
\begin{equation}
U^{u,d}_R = \left( \begin{array}{cc}
 \mathbf{1}_{3 \times 3} & \frac{y_{U,D}^{\dagger}v }{\sqrt 2 M_Q} \\
  -\frac{y_{U,D v}}{\sqrt 2 M_Q} & 1
 \end{array} \right) + \mathcal{O}\left(\frac{v^2}{M_Q^2}\right), \label{ur}
\end{equation}
so the right handed charged currents are suppressed with $V^R_{ll} \sim \mathcal{O}(v^2/M_Q^2)$ and $V^R_{lh} \sim \mathcal{O}(v/M_Q)$, where $l,h$ stand for light three generations and the extra heavy quarks, respectively.

Turning to the left handed sector, we can choose e.g. a basis of left handed quarks where $y_u$ is diagonal and real and use a unitary transformation to diagonalize $y_d$. Then we proceed in a similar way as before, and we obtain that, to first nonvanishing order in $v/M_Q$, $K_L^{u,d}$ from \eqref{rot} equal those matrices (no corrections to unity or unitarity). Combining the up and down rotations, the $3 \times 3$ upper left submatrix of $V^L$ takes the form 
\begin{equation}
V^L_{ij} = (K^u_L)_{ik} (K^{d}_L)^{ \ast}_{jk} + (R^u_L)_i (R^d_L)_j^{ \ast}\,, \quad i,j,k=1,2,3.
\end{equation}
In order to obtain the matrix elements, one has to solve the equations $\mathcal{M}^{u,d} \mathcal{M}^{u,d\dagger} U^{u,d}_L = U^{u,d}_L D_{u,d}^2$, where $D_{q} = \textrm{diag}(m_q,m_{q'})$ are the diagonal mass matrices. In the process we require that the entries in  $U^{u,d}_{L}$ mixing chiral and vector-like components ($R^{u,d}_{L}$ and $S^{u,d}_{L}$) are smaller than the remaining ones ($K^{u,d}_{L}$ and $T^{u,d}_{L}$) in terms of $v/M_Q$ scaling (similar to the case of $U^{u,d}_{R}$). Also, we assume that the corrections to the masses are small enough so that $m_q, m_{q'}$ are of order $v$ and $M_Q$, respectively. Consequently the equations for $R_L^{u,d}$ read
\begin{equation}
v^2 (y_{u,d} y_{u,d}^{\dagger} R_L^{u,d} + y_{u,d} y_{U,D}^{\dagger} T_L^{u,d}) = 2 m_{u',d'}^2 R_L^{u,d}.
\end{equation}
With the mentioned assumptions, one can neglect the first term in the above equation, obtaining $R^{u,d}_L = v^2 y_{u,d} y_{U,D}^{\dagger}/2 m_{u',d'}^2$. Thus, $R^{u,d}_L$ are one order higher than $R^{u,d}_R$ (see \ref{ur}), the off-diagonal elements in the fourth row and column are of higher order as well, which makes $K_L^d$ a unitary $3 \times 3$ matrix to second order in $v/M_Q$.
Consequently all right handed contributions, as well as corrections in the left handed sector are doubly suppressed in the loops of $B_s \to \mu^+ \mu^-$ and we can deduce a conservative bound on the FCNC's by neglecting the afore mentioned corrections, as we have done it in the vector-like singlet quark case. In fact, with the approximations made, the upper bound on $|Y_{bs}|$ is the same as the one on $|X_{bs}|$ in the down-type singlet case.

%%%%%%%%%%%%%%%%%%%%%%%%%%%%%%%%%%%%%%%%
%
\section{Constraining $Z\to q\bar q$}
\label{sec:Z}
%
%%%%%%%%%%%%%%%%%%%%%%%%%%%%%%%%%%%%%%%%

The appearance of Z-mediated FCNC's is generically connected to modifications
of diagonal $Zf\bar{f}$ couplings. In the quark sector, such effects
can be probed by direct measurements of the hadronic $Z$ decay width,
its heavy flavor decays, such as $Z\to b\bar{b}$, but also at low
energies by e.g. atomic parity violation (APV) measurements. In general we can parametrize the chiral $Zq\bar{q}$ couplings in terms of ${G}_{M}^{q}$ ($M=L,R$) as the coefficients multiplying $\ -\frac{g}{\textrm{cos}\theta_{W}}(\overline{q}_{M}\gamma^{\mu}q_{M})Z_{\mu}$ and $\delta {G}_{M}^{q}={G}_{M}^{q}-({G}_{M}^{q})_{SM}$ where $({G}_{M}^{q})_{SM}$ are given in ref.~\cite{Haisch:2011up}. In our scenarios $\delta {G}_{L}^{q} = \delta X^{u}_{qq}/2$ or $\delta {G}_{L}^{q} =- \delta X^{d}_{qq}/2$, and $\delta {G}_{R}^{q} = Y^{u}_{qq}/2$ or $\delta {G}_{R}^{q} = -Y^{d}_{qq}/2$\,.

Stringent constraints on $X^u_{uu}$ and $X^d_{dd}$ can be derived
from APV measurements in ${}^{133}{\rm Cs}$. The tree level modification
of the $Z$ boson couplings to first generation quarks will modify
the nuclear weak charge as~\cite{Gresham:2012wc} 
\begin{equation}
\delta Q_{W}(Z,N)=2(2Z+N)(\delta{G}_{L}^{u}+\delta{G}_{R}^{u})+2(Z+2N)(\delta{G}_{L}^{d}+\delta{G}_{R}^{d})\,,
\end{equation}
where the measured value deviates from the SM expectation by $1.5\,\sigma$~\cite{Dzuba:2012kx}
\begin{equation}
Q_{W}-Q_{W}^{SM}\equiv\delta Q_{W}=0.65(43).
\end{equation}
For the singlet up-like vector-like quark, this constrains $ \delta X_{uu}^{u}=0.0035(23)$,
while for the singlet down-like vector-like quark $\delta  X_{dd}^{d}=-0.0031(20)$.
In the doublet case, one has two independent variables $\delta G_{R}^{u}$
and $\delta G_{R}^{d}$ at the tree level, and only a certain linear combination
can be constrained, namely $ Y_{uu}^{u}-1.12 Y_{dd}^{d}=0.0035(23)$.

\begin{table}
\begin{centering}
\begin{tabular}{|c|c|c|c|}
\hline 
Observable & Measured value & SM prediction & Reference\tabularnewline
\hline 
\hline 
$R_{b}$ & $0.21629(66)$ & $0.21474(3)$ & \cite{ALEPH:2005ab,Batell:2012ca} \tabularnewline
\hline 
$A_{b}$ & $0.923(20)$ & $0.9347(1)$ & \cite{pdg,ALEPH:2005ab} 
 \tabularnewline
\hline 
$A_{FB}^{b}$ & $0.0992(16)$ & $0.1032(6)$ & \cite{ALEPH:2005ab,Batell:2012ca}\tabularnewline
\hline 
$\sigma_{had}[\textrm{nb}]$ & $41.541(37)$ & $41.477(9)$ &  \cite{pdg,ALEPH:2005ab} \tabularnewline
\hline 
$\Gamma_{Z}[\textrm{MeV}]$ & $2495.2(2.3)$ & $2495.5(1.0)$ &  \cite{pdg,ALEPH:2005ab} \tabularnewline
\hline 
$R_{c}$ & $0.1721(30)$ & $0.17227(4)$ &  \cite{pdg,ALEPH:2005ab} \tabularnewline
\hline 
$A_{c}$ & $0.670(27)$ & $0.6680(4)$ &  \cite{pdg,ALEPH:2005ab} \tabularnewline
\hline 
$A_{FB}^{c}$ & $0.0707(35)$ & $0.0739(5)$ &  \cite{pdg,ALEPH:2005ab} \tabularnewline
\hline 
\end{tabular}
\par\end{centering}

\caption{Z pole observables used in the analysis and their SM predictions.
Correlations between observables are neglected. \label{tab:Zpole}}
\end{table}
Additional constraints can be derived from direct $Z\to q\bar{q}$ measurements
presented in Table~\ref{tab:Zpole}. In particular, the total Z decay width is given
by
\begin{equation}
\Gamma_{Z}=\Gamma_{\textrm{inv}}+\Gamma_{\textrm{lep}}+\Gamma_{\textrm{had}}.
\end{equation}
It incorporates the decays to leptons, where $\Gamma_{\textrm{lep}}=251.7$
MeV in the SM, decays into invisible particles (neutrinos), being
$\Gamma_{\textrm{inv}}=501.6$ MeV in the SM, and the decays into
hadrons. The hadronic width is the sum over the decays into all kinematically
accessible quarks, that is all SM quarks but the top, $\Gamma_{\textrm{had}}=\sum_{q}\ \Gamma_{q}$.
The partial $Z$-decay width to light quarks is given by
\[
\Gamma_{q}\equiv\Gamma(Z\to q\bar{q})=N_{C}\frac{G_{F}M_{Z}^{3}}{6\sqrt{2}\pi}\left(R_{V}^{q}\left|{G}_{L}^{q}+{G}_{R}^{q}\right|^{2}+R_{A}^{q}\left|{G}_{L}^{q}-{G}_{R}^{q}\right|^{2}\right)+\Delta_{EW,QCD}^{q},
\]
with radiator factors $R_{V}^{q}$ and $R_{A}^{q}$ and non-factorizable
radiative correction parameters $\Delta_{EW,QCD}^{q}$ given in~\cite{Haisch:2011up}. The fractions of hadronic $Z$ decays involving $b$ quark
pairs and $c$ quark pairs are defined as
\begin{equation}
R_{b}=\frac{\Gamma_{b}}{\Gamma_{\textrm{had}}},\;\; R_{c}=\frac{\Gamma_{c}}{\Gamma_{\textrm{had}}},
\end{equation}
respectively. The associated bottom and charm quark left-right asymmetries
($A_{b}$ and $A_{c}$), and forward-backward asymmetries ($A_{FB}^{b}$
and $A_{FB}^{c}$) can be written as 
\begin{equation}
A_{f}=\frac{2\sqrt{1-4z_{f}}\ \frac{{G}_{L}^{f}+{G}_{R}^{f}}{{G}_{L}^{f}-{G}_{R}^{f}}}{1-4z_{f}+(1+2z_{f})\left(\frac{{G}_{L}^{f}+{G}_{R}^{f}}{{G}_{L}^{f}-{G}_{R}^{f}}\right)^{2}},\;\; A_{FB}^{f}=\frac{3}{4}A_{e}A_{f} ,
\end{equation}
where $f=b,c$ and $z_{f}=m_{f}^{2}(m_{Z})/m_{Z}^{2}$. The electron asymmetry
parameter is fixed to its SM value, $(A_{e})_{SM}=0.1464$.

\begin{figure}
\begin{centering}
\includegraphics[scale=1.2]{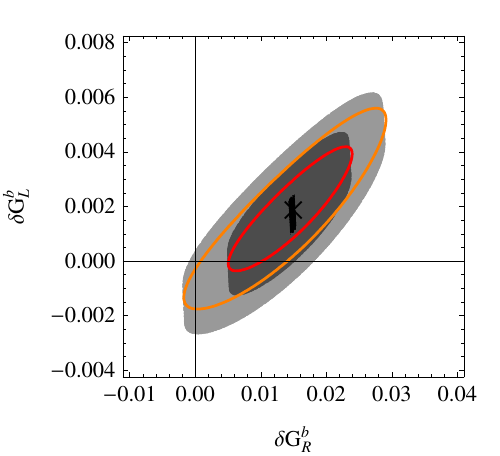}$\;\;\;\;$\includegraphics[scale=1.2]{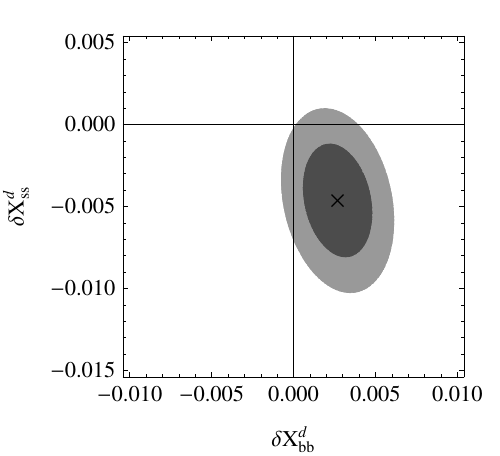}
\par\end{centering}
\caption{\textbf{Left:} Fit of Z-pole data taking $\delta G_{L,R}^q$ for $q=u,d,c,s,b$ as fitting parameters. Best fit point (cross), $1\,\sigma$ (dark gray) and $2\,\sigma$ (light gray) regions are shown in the $(\delta G_R^b,\delta G_L^b)$ plane after marginalizing over the other parameters. The results of the fit for fixed $\delta G_{L,R}^{u,d,c,s}=0$ are given by the red contour ($1\,\sigma$ region) and the orange contour ($2\,\sigma$ region). \textbf{Right:}  The fit of Z-pole data in the model with a singlet down-like vector-like quark. Best fit point (cross), $1\,\sigma$ (dark gray) and $2\,\sigma$ (light gray) regions are shown in $(\delta X_{bb}^d,\delta X_{ss}^d)$ plane.    \label{fig:Zfit}}
\end{figure}

Another interesting quantity is the hadronic $e^{+}e^{-}$ cross section
at the $Z$ pole ($\sigma_{\textrm{had}}$). It can be written as
\begin{equation}
\sigma_{\textrm{had}}=\frac{12\pi}{M_{Z}^{2}}\frac{\Gamma_{e}\Gamma_{\textrm{had}}}{\Gamma_{Z}^{2}},
\end{equation}
where $\Gamma_{e}=84.005\,$MeV and $M_{z}=91.1876\,$GeV. We perform a $\chi^{2}$
fit of the data presented in Table~\ref{tab:Zpole} in terms of
$\delta G_{M}^{q}$ as fitting parameters. The results are presented in the $(\delta{G}_{R}^{b},\delta{G}_{L}^{b})$ plane, after marginalizing over all other variables, in Fig.~\ref{fig:Zfit} left. 
At the best fit point
$\chi_{min}^{2}=3.7$, and the most important observables in
the fit are $R_{b}$, $A_{FB}^{b}$ and $\sigma_{had}$, which contribute
to $\chi_{SM}^{2}-\chi_{min}^{2}$ by $5.6$, $4.9$ and $1.3$, respectively.
In the SM reference scenario $\chi_{SM}^{2}=14.9$, corresponding to a p-value
of $0.06$. The largest contributions to $\chi_{SM}^{2}$
come from $R_{b}$, $A_{FB}^{b}$ and $\sigma_{had}$ and are $5.6$,
$5.2$ and $2.8$ respectively. We also perform a $\chi^{2}$ fit only with $\delta G_{R}^{b}$ and $\delta G_{L}^{b}$
while putting other parameters to zero. In this case $\chi^{2}_{min}=7.8$ with p-value $0.25$. The corresponding $1\,\sigma$ and $2\,\sigma$
regions are presented in the left Fig.~\ref{fig:Zfit} by red and orange curves, respectively. As expected, the data is mainly sensitive to bottom quark couplings leading to model independent bounds of $\delta{G}_{L}^{b}=0.002(2)$ and $\delta{G}_{R}^{b}=0.015(6)$.

Now, we turn to specific models. In the singlet up-type vector-like
quark model, tree level modification of $\delta{G}_{L}^{u}$
and $\delta{G}_{L}^{c}$ is possible. Taking into account the severe constraint on $\delta X_{uu}^{u}=-0.0001(6)$
from CKM unitarity, $\delta{G}_{L}^{u}$ has no significant
influence in the fit. Therefore, we use the data to extract the best current
constraint on $\delta X_{cc}^{u}=-0.0020(13)$. The constraint comes essentially
 from three observables: $R_{b}$, $\sigma_{had}$ and $\Gamma_{z}$, which are more constraining than direct $Z\to c\bar c$ measurements $R_c$ and $A^c_{FB}$.
We calculate the individual contribution to $\Delta\chi^{2}$ from each
observable in the points which are $\pm1\,\sigma$ away from the best
fit point. Contributions to $\Delta\chi_{-\sigma}^{2}$ from $R_{b}$,
$\sigma_{had}$, $\Gamma_{z}$ , $R_{c}$ and $A_{FB}^{c}$ are $ $$-1.1$,
$-0.7$, $2.5$, $0.4$ and $-0.1$, respectively, while contributions
to $\Delta\chi_{+\sigma}^{2}$ from $R_{b}$, $\sigma_{had}$, $\Gamma_{z}$,
$R_{c}$ and $A_{FB}^{c}$ are $ $$1.4$, $1.1$, $-1.4$, $-0.2$
and $0.1$, respectively.

In the singlet down-type vector-like quark model, tree level modifications
of $\delta{G}_{L}^{d}$ , $\delta{G}_{L}^{s}$ and
$\delta{G}_{L}^{b}$ are possible. APV puts a strong constraint on $\delta X_{dd}^{d}$,
so that the $Z$ data can be  fitted with only two parameters, $\delta  X_{ss}^{d}$ and $\delta  X_{bb}^{d}$.
Results are presented in Fig.~\ref{fig:Zfit} right. The best fit point now corresponds
to $( \delta X_{bb}^{d}, \delta X_{ss}^{d})=(0.0027,-0.0046)$
where $\chi_{min}^{2}=8.7$. The fit is statistically better than in
the SM with a p-value of $0.2$. The most relevant observable
in the fit is $R_{b}$ and its contribution to $\chi_{SM}^{2}-\chi_{min}^{2}$
is $5.6$, while the contribution of $\sigma_{had}$ is $1.0$. Analyzing
one-dimensional $\chi^{2}$ functions for each observable, we find
$1\,\sigma$ preferred regions to be $ \delta X_{ss}^{d}=-0.005(2)$
and $\delta  X_{bb}^{d}=0.0027(15)$. Finally, including $ \delta X_{dd}^{d}$
in the fit and taking into account the APV constraint, the preferred region
for  $ \delta X_{ss}^{d}$ is reduced to $\delta  X_{ss}^{d} =-0.002(3)$,
while the preferred region for $ \delta X_{bb}^{d}$ is unaffected.

In the doublet vector-like quark model, tree level modification of
all right handed couplings of light quarks with Z boson is possible.
Therefore, we fit Z-pole observables together with the APV constraint on
$ Y_{uu}^{u}-1.12 Y_{dd}^{d}$ with five parameters $ Y_{uu}^{u}$,
$ Y_{cc}^{u}$, $ Y_{dd}^{d}$, $ Y_{ss}^{d}$ and
$ Y_{bb}^{d}$. Marginalizing over four parameters, we get the
preferred range for the remaining one. Modification of charm and bottom
quark couplings can be constrained to a percent level, $ Y_{bb}^{d}=-0.018(6)$
and $ Y_{cc}^{u}=-0.003(9)$. Negative value for $ Y_{bb}^{d}$
is mainly driven by $R_{b}$ and $A_{FB}^{b}$, which contribute to
$\chi_{SM}^{2}-\chi_{min}^{2}$ by $5.5$ and $3.5$, respectively.
In the case of the light quark couplings, observables given in the
table can not distinguish between different light flavors, giving very
poor constraints on one coupling after marginalizing over others.
Therefore, we include new observables into the fit which are poorly measured
but can distinguish between light quark flavours, namely asymmetries associated with the strange
quark, with experimental values $(A^{s})_{exp}=0.895(91)$ and $(A_{FB}^{s})_{exp}=0.0976(114)$~\cite{pdg}. The constraints we get are rather mild, since the experimental
values for $A_{s}$ and $A_{FB}^{s}$ have large experimental uncertainties.
We obtain the following bounds on $ Y_{uu}^{u}=0.035(40)$, $ Y_{dd}^{d}=0.030(35)$
and $ Y_{ss}^{d}=-0.05_{-0.06}^{+0.08}$.

\begin{figure}
\begin{centering}
\includegraphics{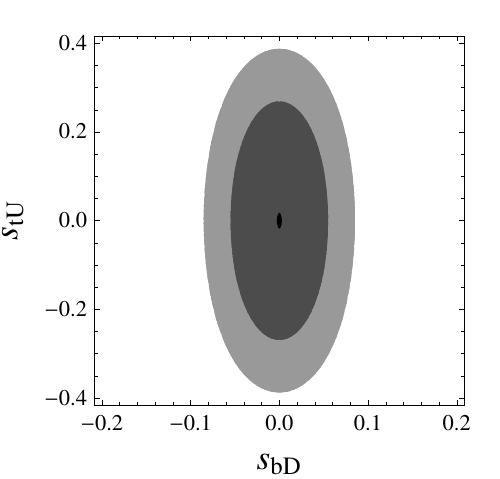}
\par\end{centering}
\caption{Fit of Z-pole data in the model with a doublet vector-like quark mixing predominantly
with the third generation. Best fit point, $1\,\sigma$ (dark gray) and $2\,\sigma$ (light gray) regions are shown in the $(s_{bD},s_{tU})$ plane.\label{fig:tb}}
\end{figure}

Finally, in the model with one vector-like quark doublet mixing predominantly
with the third generation, modification of $\delta {G}_{R}^{b}$ is
induced at tree level $\delta {G}_{R}^{b}=-(1/2)s^2_{bD}$,
while modification of $\delta {G}_{L}^{b}$ is induced at one-loop level
\begin{equation}
\delta {G}_{L}^{b}=\frac{g^{2}}{64\pi^{2}}s^2_{tU}\left(\frac{f_{1}(x,x')}{r}+f_{3}(x,x')\right),
\end{equation}
where $x\equiv m_{t}^{2}/m_{W}^{2}$, $x'\equiv m_{u'}^{2}/m_{W}^{2}$
and $r\equiv m_{u'}^{2}/m_{t}^{2}$, and $f_{1}$ and $f_{3}$ are
loop functions given in refs.~\cite{Sally,Bamert}. The above expression
is given in the $x,x'\gg1$  limit and for small mixing angles. In our numerical evaluation we use the full one-loop expressions and do not assume small mixing angles, even though we note that the above approximations are  fairly good. Using the available $Z$-pole data to fit $s_{bD}$ and $s_{tU}$
parameters, we present the results in the $(s_{bD},s_{tU})$
plane in Fig.$\,$\ref{fig:tb} for fixed $m_{u'}=800\,$GeV. 
We have checked, however,
that the results are mostly insensitive to the precise value of the $u'$ mass in the interesting region ($500\,{\rm GeV}<m_{u'}<1500$\,GeV).  Marginalizing over $s_{bD}$, we obtain $\left|s_{tU}\right|<0.35$
  at $95\%$ C.L.\,.

%%%%%%%%%%%%%%%%%%%%%%%%%%%%%%%%%%%%%%%%
%
\section{Doublet vector-like quark contributions to $\rho$ parameter in presence of $1/\Lambda$ corrections}
\label{sec:rhoDoublet}
%
%%%%%%%%%%%%%%%%%%%%%%%%%%%%%%%%%%%%%%%%

In the renormalizable doublet vector-like quark model, the divergences in the loop calculation of the $\rho$ parameter vanish only after imposing the connection between masses and mixing angles in the up- and down-quark sectors namely
\begin{align}
M_D &= M_U \equiv M_Q \label{equ}\,,
\end{align}
with
\begin{align}
M_D^2 &=  c_{bD}^2 m^2_{d'} + s_{bD}^2 m_{b}^2 \,,\nonumber\\
M_U^2 &=  c_{tU}^2 m^2_{u'} + s_{tU}^2 m_{t}^2\,. \label{UD}
\end{align}
Relation \ref{equ} no longer holds after the inclusion of non-renormalizable operators. However, one can still relate the parameters in the up and down sector through the identities (see Sec.~\ref{sec:V})
\begin{equation}
M_{U,D} \equiv \sqrt{\bar M_{U,D}^2 + v^4 ( (c_1^-)^i (c_1^-)^*_i)/4\Lambda^2  }\,,\qquad \bar M_{U,D} \equiv M_{Q}  \pm v^2 c_2^-/2 \Lambda\,.
\end{equation} 
Expanding to $\mathcal{O}(1/\Lambda)$, the divergences in $\rho$ again cancel. %For consistency reasons, the same expansion has to be performed in the finite part. 
Building upon known oblique corrections from vector-like quarks in the renormalizable model~\cite{Lavoura:1992np}, the new physics contribution to the $\rho$ parameter including leading dimension 5 non-renormalizable operators can then be written as
\begin{align}
\Delta\rho & = \frac{\alpha N_C}{16 \pi s_{W}^2} \Bigg\lbrace \sum_{i=t,u',j=b,d'} \left[ \left(|\tilde V^L_{ij}|^2+|\tilde V^R_{ij}|^2\right) g_1(x_i,x_j) + 2 \textrm{Re}\left(\tilde V^L_{ij}\tilde V^{R\ast}_{ij}\right) g_2(x_i,x_j) \right] \nonumber\\ 
& - s_{bD}^2 c_{bD}^2 g_1(x_{d'},x_b) - s_{tU}^2 c_{tU}^2 g_1(x_{u'},x_t)  - g_1(x_t,x_b)  + g_{\textrm{nr}} \Bigg\rbrace\,,
\end{align}
%also there are no LH FCNC's, so no g2 in NC's
 where $x_i \equiv m_i^2 / m_W^2$. The corresponding mixing matrices are defined as $(\tilde V^{L})_{ij} \equiv  (\tilde U_u^{L})_{ik} (\tilde U_d^{L})^*_{jk}$ and $(\tilde V^{R})_{ij}\equiv  (\tilde U_u^{R})_{i2} (\tilde U_d^{R})_{j2}^*$, where 
\beq
\tilde U^R_{u,d} = \left( \begin{array}{ccc}
c_{tU,bD} & s_{tU,bD} \\
-s_{tU,bD} & c_{tU,bD} \\
\end{array}\right)\,, \quad \tilde U^L_{u,d} = \left( \begin{array}{ccc}
c'_{tU,bD} & s'_{tU,bD} \\
-s'_{tU,bD} & c'_{tU,bD} \\
\end{array}\right)\,.
\eeq
The relevant loop functions are given by
\begin{align}
g_1(x_1,x_2) & \equiv x_1 + x_2 - \frac{2 x_1 x_2}{x_1-x_2} \ln \frac{x_1}{x_2}\,, \\
g_2(x_1,x_2) & \equiv 2 \sqrt{x_1x_2} \left( \frac{x_1 + x_2}{x_1-x_2} \ln \frac{x_1}{x_2} - 2 \right),
\end{align}
while the new term which incorporates explicit effects due to $\mathcal O(1/\Lambda)$ non-renormalizable mass-splitting operators reads
\begin{equation}
g_{\textrm{nr}} = 4 \delta \left\lbrace s_{bD}^2 x_b \ln \frac{x_b}{x_{d'}} - s_{tU}^2 x_t \ln \frac{x_t}{x_{u'}} + x_U \ln \frac{x_{d'}}{x_{u'}} \right\rbrace ,
\end{equation}
with $\delta \equiv (M_U - M_D)/ M_U$ and $M_{U,D}$ given in \ref{UD} (note that $x_U$ refers to $M_U^2/m_W^2$).\\

%%%%%%%%%%%%%%%%%%%%%%%%%%%%%%%%%%%%%%%%
%%%%%%%%%%%%%%%%%%%%%%%%%%%%%%%%%%%%%%%%
%%%%%%%%%%%%%%%%%%%%%%%%%%%%%%%%%%%%%%%%


\begin{thebibliography}{99}  

\bibitem{Aad:2012tfa}
  G.~Aad {\it et al.}  [ATLAS Collaboration],
  %``Observation of a new particle in the search for the Standard Model Higgs boson with the ATLAS detector at the LHC,''
  Phys.\ Lett.\ B {\bf 716} (2012) 1
  [arXiv:1207.7214 [hep-ex]];
  %%CITATION = ARXIV:1207.7214;%%
  %897 citations counted in INSPIRE as of 09 Apr 2013
  S.~Chatrchyan {\it et al.}  [CMS Collaboration],
  %``Observation of a new boson at a mass of 125 GeV with the CMS experiment at the LHC,''
  Phys.\ Lett.\ B {\bf 716} (2012) 30
  [arXiv:1207.7235 [hep-ex]].
  %%CITATION = ARXIV:1207.7235;%%
  %888 citations counted in INSPIRE as of 09 Apr 2013

\bibitem{Papucci:2011wy} 
  M.~Papucci, J.~T.~Ruderman and A.~Weiler,
  %``Natural SUSY Endures,''
  JHEP {\bf 1209}, 035 (2012)
  [arXiv:1110.6926 [hep-ph]].
  %%CITATION = ARXIV:1110.6926;%%
  %150 citations counted in INSPIRE as of 09 Apr 2013
 
 \bibitem{ArkaniHamed:2001nc} 
  N.~Arkani-Hamed, A.~G.~Cohen and H.~Georgi,
  %``Electroweak symmetry breaking from dimensional deconstruction,''
  Phys.\ Lett.\ B {\bf 513}, 232 (2001)
  [hep-ph/0105239];
  %%CITATION = HEP-PH/0105239;%%
  %904 citations counted in INSPIRE as of 09 Apr 2013
  N.~Arkani-Hamed, A.~G.~Cohen, T.~Gregoire and J.~G.~Wacker,
  %``Phenomenology of electroweak symmetry breaking from theory space,''
  JHEP {\bf 0208}, 020 (2002)
  [hep-ph/0202089];
  %%CITATION = HEP-PH/0202089;%%
  %219 citations counted in INSPIRE as of 09 Apr 2013
  N.~Arkani-Hamed, A.~G.~Cohen, E.~Katz, A.~E.~Nelson, T.~Gregoire and J.~G.~Wacker,
  %``The Minimal moose for a little Higgs,''
  JHEP {\bf 0208}, 021 (2002)
  [hep-ph/0206020];
  %%CITATION = HEP-PH/0206020;%%
  %436 citations counted in INSPIRE as of 09 Apr 2013
  M.~Perelstein, M.~E.~Peskin and A.~Pierce,
  %``Top quarks and electroweak symmetry breaking in little Higgs models,''
  Phys.\ Rev.\ D {\bf 69}, 075002 (2004)
  [hep-ph/0310039];
  %%CITATION = HEP-PH/0310039;%%
  %164 citations counted in INSPIRE as of 12 Apr 2013
  T.~Han, H.~E.~Logan, B.~McElrath and L.~-T.~Wang,
  %``Phenomenology of the little Higgs model,''
  Phys.\ Rev.\ D {\bf 67}, 095004 (2003)
  [hep-ph/0301040].
  %%CITATION = HEP-PH/0301040;%%
  %407 citations counted in INSPIRE as of 12 Apr 2013

  
  \bibitem{Panico:2012uw}
  R.~Contino, L.~Da Rold and A.~Pomarol,
  %``Light custodians in natural composite Higgs models,''
  Phys.\ Rev.\ D {\bf 75}, 055014 (2007)
  [hep-ph/0612048];
  %%CITATION = HEP-PH/0612048;%%
  %186 citations counted in INSPIRE as of 12 Apr 2013
  C.~Anastasiou, E.~Furlan and J.~Santiago,
  %``Realistic Composite Higgs Models,''
  Phys.\ Rev.\ D {\bf 79}, 075003 (2009)
  [arXiv:0901.2117 [hep-ph]];
  %%CITATION = ARXIV:0901.2117;%%
  %35 citations counted in INSPIRE as of 12 Apr 2013 
  O.~Matsedonskyi, G.~Panico and A.~Wulzer,
  %``Light Top Partners for a Light Composite Higgs,''
  JHEP {\bf 1301}, 164 (2013)
  [arXiv:1204.6333 [hep-ph]];
  %%CITATION = ARXIV:1204.6333;%%
  %23 citations counted in INSPIRE as of 12 Apr 2013
  G.~Panico, M.~Redi, A.~Tesi and A.~Wulzer,
  %``On the Tuning and the Mass of the Composite Higgs,''
  JHEP {\bf 1303}, 051 (2013)
  [arXiv:1210.7114 [hep-ph]];
  %%CITATION = ARXIV:1210.7114;%%
  %10 citations counted in INSPIRE as of 11 Apr 2013
  M.~Redi and A.~Tesi,
  %``Implications of a Light Higgs in Composite Models,''
  JHEP {\bf 1210}, 166 (2012)
  [arXiv:1205.0232 [hep-ph]];
  %%CITATION = ARXIV:1205.0232;%%
  %23 citations counted in INSPIRE as of 12 Apr 2013
  L.~Vecchi,
  %``The Natural Composite Higgs,''
  arXiv:1304.4579 [hep-ph].
  
  \bibitem{Giudice:2007fh} 
  G.~F.~Giudice, C.~Grojean, A.~Pomarol and R.~Rattazzi,
  %``The Strongly-Interacting Light Higgs,''
  JHEP {\bf 0706}, 045 (2007)
  [hep-ph/0703164].
  %%CITATION = HEP-PH/0703164;%%
  %231 citations counted in INSPIRE as of 09 Apr 2013
  
  \bibitem{Contino:2013kra} 
  R.~Contino, M.~Ghezzi, C.~Grojean, M.~Muhlleitner and M.~Spira,
  %``Effective Lagrangian for a light Higgs-like scalar,''
  arXiv:1303.3876 [hep-ph].
  %%CITATION = ARXIV:1303.3876;%%
  %3 citations counted in INSPIRE as of 09 Apr 2013
  
  \bibitem{Azatov:2011qy} 
  A.~Azatov and J.~Galloway,
  %``Light Custodians and Higgs Physics in Composite Models,''
  Phys.\ Rev.\ D {\bf 85}, 055013 (2012)
  [arXiv:1110.5646 [hep-ph]];
  %%CITATION = ARXIV:1110.5646;%%
  %30 citations counted in INSPIRE as of 11 Apr 2013
  J.~Berger, J.~Hubisz and M.~Perelstein,
  %``A Fermionic Top Partner: Naturalness and the LHC,''
  JHEP {\bf 1207}, 016 (2012)
  [arXiv:1205.0013 [hep-ph]];
  %%CITATION = ARXIV:1205.0013;%%
  %16 citations counted in INSPIRE as of 11 Apr 2013
  A.~Carmona and F.~Goertz,
  %``Custodial Leptons and Higgs Decays,''
  arXiv:1301.5856 [hep-ph].
  
  \bibitem{AguilarSaavedra:2002kr} 
  J.~A.~Aguilar-Saavedra,
  %``Effects of mixing with quark singlets,''
  Phys.\ Rev.\ D {\bf 67}, 035003 (2003)
  [Erratum-ibid.\ D {\bf 69}, 099901 (2004)]
  [hep-ph/0210112].
  %%CITATION = HEP-PH/0210112;%%
  %98 citations counted in INSPIRE as of 09 Apr 2013
  
    \bibitem{Dowdall:2013rya} 
  R.~J.~Dowdall, C.~T.~H.~Davies, G.~P.~Lepage and C.~McNeile,
  %``Vus from pi and K decay constants in full lattice QCD with physical u, d, s and c quarks,''
  arXiv:1303.1670 [hep-lat].
  %%CITATION = ARXIV:1303.1670;%%


  \bibitem{pdg} 
  J.~Beringer {\it et al.}  [Particle Data Group Collaboration],
  %``Review of Particle Physics (RPP),''
  Phys.\ Rev.\ D {\bf 86}, 010001 (2012).
  %%CITATION = PHRVA,D86,010001;%%
  %1160 citations counted in INSPIRE as of 09 Apr 2013
  
  \bibitem{Buras:2010pz} 
  A.~J.~Buras, K.~Gemmler and G.~Isidori,
  %``Quark flavour mixing with right-handed currents: an effective theory approach,''
  Nucl.\ Phys.\ B {\bf 843}, 107 (2011)
  [arXiv:1007.1993 [hep-ph]].
  %%CITATION = ARXIV:1007.1993;%%
  %51 citations counted in INSPIRE as of 09 Apr 2013
  
  \bibitem{Golowich:2007ka} 
  E.~Golowich, J.~Hewett, S.~Pakvasa and A.~A.~Petrov,
  %``Implications of $D^0$ - $\bar{D}^0$ Mixing for New Physics,''
  Phys.\ Rev.\ D {\bf 76}, 095009 (2007)
  [arXiv:0705.3650 [hep-ph]].
  %%CITATION = ARXIV:0705.3650;%%
  %147 citations counted in INSPIRE as of 09 Apr 2013
  
    \bibitem{hfag}
  Y.~Amhis {\it et al.}  [Heavy Flavor Averaging Group Collaboration],
  %``Averages of B-Hadron, C-Hadron, and tau-lepton properties as of early 2012,''
  arXiv:1207.1158 [hep-ex].
  %%CITATION = ARXIV:1207.1158;%%
  %136 citations counted in INSPIRE as of 09 Apr 2013

  \bibitem{dipole}
  B.~Grzadkowski and M.~Misiak,
  %``Anomalous Wtb coupling effects in the weak radiative B-meson decay,''
  Phys.\ Rev.\ D {\bf 78}, 077501 (2008)
  [Erratum-ibid.\ D {\bf 84}, 059903 (2011)]
  [arXiv:0802.1413 [hep-ph]];
  %%CITATION = ARXIV:0802.1413;%%
  %66 citations counted in INSPIRE as of 08 Jun 2013
  J.~Drobnak, S.~Fajfer and J.~F.~Kamenik,
  %``Interplay of t --> b W Decay and B_q Meson Mixing in Minimal Flavor Violating Models,''
  Phys.\ Lett.\ B {\bf 701}, 234 (2011)
  [arXiv:1102.4347 [hep-ph]];
  %%CITATION = ARXIV:1102.4347;%%
  %17 citations counted in INSPIRE as of 08 Jun 2013
  J.~F.~Kamenik, M.~Papucci and A.~Weiler,
  %``Constraining the dipole moments of the top quark,''
  Phys.\ Rev.\ D {\bf 85}, 071501 (2012)
  [arXiv:1107.3143 [hep-ph]];
  %%CITATION = ARXIV:1107.3143;%%
  %19 citations counted in INSPIRE as of 08 Jun 2013
J.~Drobnak, S.~Fajfer and J.~F.~Kamenik,
  %``Probing anomalous tWb interactions with rare B decays,''
  Nucl.\ Phys.\ B {\bf 855}, 82 (2012)
  [arXiv:1109.2357 [hep-ph]];
  %%CITATION = ARXIV:1109.2357;%%
  %21 citations counted in INSPIRE as of 08 Jun 2013  
  C.~Zhang, N.~Greiner and S.~Willenbrock,
  %``Constraints on Non-standard Top Quark Couplings,''
  Phys.\ Rev.\ D {\bf 86}, 014024 (2012)
  [arXiv:1201.6670 [hep-ph]].
  %%CITATION = ARXIV:1201.6670;%%
  %14 citations counted in INSPIRE as of 08 Jun 2013
  
  
  \bibitem{Cheng:1987rs} 
  T.~P.~Cheng and M.~Sher,
  %``Mass Matrix Ansatz and Flavor Nonconservation in Models with Multiple Higgs Doublets,''
  Phys.\ Rev.\ D {\bf 35}, 3484 (1987).
  %%CITATION = PHRVA,D35,3484;%%
  %371 citations counted in INSPIRE as of 09 Apr 2013
  
  \bibitem{Harnik:2012pb} 
  G.~Blankenburg, J.~Ellis and G.~Isidori,
  %``Flavour-Changing Decays of a 125 GeV Higgs-like Particle,''
  Phys.\ Lett.\ B {\bf 712}, 386 (2012)
  [arXiv:1202.5704 [hep-ph]];
  %%CITATION = ARXIV:1202.5704;%%
  %13 citations counted in INSPIRE as of 12 Apr 2013
  R.~Harnik, J.~Kopp and J.~Zupan,
  %``Flavor Violating Higgs Decays,''
  JHEP {\bf 1303}, 026 (2013)
  [arXiv:1209.1397 [hep-ph]].
  %%CITATION = ARXIV:1209.1397;%%
  %10 citations counted in INSPIRE as of 09 Apr 2013
  
  
  \bibitem{ATLAS-2013-034} [ATLAS Collaboration], Note ATLAS-CONF-2013-034 [http://cds.cern.ch/record/1528170]\,.

\bibitem{CMS-12-045}CMS-PAS-HIG-12-045


  \bibitem{Martin:2009iq} 
  A.~D.~Martin, W.~J.~Stirling, R.~S.~Thorne and G.~Watt,
  %``Parton distributions for the LHC,''
  Eur.\ Phys.\ J.\ C {\bf 63}, 189 (2009)
  [arXiv:0901.0002 [hep-ph]].
  %%CITATION = ARXIV:0901.0002;%%
  %1385 citations counted in INSPIRE as of 10 Apr 2013
  
  \bibitem{Ball:2013bra} 
  R.~D.~Ball, M.~Bonvini, S.~Forte, S.~Marzani and G.~Ridolfi,
  %``Higgs production in gluon fusion beyond NNLO,''
  arXiv:1303.3590 [hep-ph].
  %%CITATION = ARXIV:1303.3590;%%
  %1 citations counted in INSPIRE as of 10 Apr 2013
  
  \bibitem{Group:2012zca} 
  C.~a.~D.~C.~a.~t.~T.~N.~P.~a.~H.~W.~Group [Tevatron New Physics Higgs Working Group and CDF and D0 Collaborations],
  %``Updated Combination of CDF and D0 Searches for Standard Model Higgs Boson Production with up to 10.0 fb$^{-1}$ of Data,''
  arXiv:1207.0449 [hep-ex].
  %%CITATION = ARXIV:1207.0449;%%
  %111 citations counted in INSPIRE as of 10 Apr 2013
  
  \bibitem{Djouadi:1999rca} 
  A.~Djouadi, W.~Kilian, M.~Muhlleitner and P.~M.~Zerwas,
  %``Production of neutral Higgs boson pairs at LHC,''
  Eur.\ Phys.\ J.\ C {\bf 10}, 45 (1999)
  [hep-ph/9904287].
  %%CITATION = HEP-PH/9904287;%%
  %119 citations counted in INSPIRE as of 10 Apr 2013
  
    \bibitem{Djouadi:2005gi} 
  A.~Djouadi,
  ``The Anatomy of electro-weak symmetry breaking. I: The Higgs boson in the standard model,''
  Phys.\ Rept.\  {\bf 457}, 1 (2008)
  [hep-ph/0503172].
  %CITATION = HEP-PH/0503172;%%
  %629 citations counted in INSPIRE as of 09 Apr 2013

\bibitem{Cacciapaglia:2012wb} 
  G.~Cacciapaglia, A.~Deandrea, G.~D.~La Rochelle and J.~-B.~Flament,
  %``Higgs couplings beyond the Standard Model,''
  JHEP {\bf 1303}, 029 (2013)
  [arXiv:1210.8120 [hep-ph]];
  %%CITATION = ARXIV:1210.8120;%%
  %21 citations counted in INSPIRE as of 07 Jun 2013
  G.~Belanger, B.~Dumont, U.~Ellwanger, J.~F.~Gunion and S.~Kraml,
  %``Higgs Couplings at the End of 2012,''
  JHEP {\bf 1302}, 053 (2013)
  [arXiv:1212.5244 [hep-ph]].
  %%CITATION = ARXIV:1212.5244;%%
  %23 citations counted in INSPIRE as of 07 Jun 2013
  
\bibitem{key-1}[ATLAS Collaboration], Note ATLAS-CONF-2013-013, [http://cds.cern.ch/record/1523699].

\bibitem{key-W}[ATLAS Collaboration], Note ATLAS-CONF-2013-030, [http://cds.cern.ch/record/1527126].

\bibitem{key-3}[ATLAS Collaboration], Note ATLAS-CONF-2013-012, [http://cds.cern.ch/record/1523698].

\bibitem{key-Cw}[CMS Collaboration], Note CMS-PAS-HIG-13-003, [http://cds.cern.ch/record/1523673].

\bibitem{key-5}[CMS Collaboration], Note CMS-PAS-HIG-13-002, [http://cds.cern.ch/record/1523767].

\bibitem{key-9}Talk by Christophe Ochando, Moriond QCD and High Energy
Interactions, 14.03.2013.

\bibitem{key-tau}[CMS Collaboration], Note CMS-PAS-HIG-13-004, [http://cds.cern.ch/record/1528271];
Talk by Mingshui Chen, Rencontres de Moriond EW, 06.03.2013;
Talk by Valentina Duta, Rencontres de Moriond EW, 06.03.2013.


\bibitem{Delaunay:2013iia} 
  C.~Delaunay, C.~Grojean and G.~Perez,
  %``Modified Higgs Physics from Composite Light Flavors,''
  arXiv:1303.5701 [hep-ph].
  %%CITATION = ARXIV:1303.5701;%%
  %1 citations counted in INSPIRE as of 10 Apr 2013

\bibitem{Kearney:2012zi} 
  J.~Kearney, A.~Pierce and N.~Weiner,
  %``Vectorlike Fermions and Higgs Couplings,''
  Phys.\ Rev.\ D {\bf 86}, 113005 (2012)
  [arXiv:1207.7062 [hep-ph]];
  %%CITATION = ARXIV:1207.7062;%%
  %29 citations counted in INSPIRE as of 12 Apr 2013
  G.~Moreau,
  %``Constraining extra-fermion(s) from the Higgs boson data,''
  Phys.\ Rev.\ D {\bf 87}, 015027 (2013)
  [arXiv:1210.3977 [hep-ph]].
  %%CITATION = ARXIV:1210.3977;%%
  %18 citations counted in INSPIRE as of 12 Apr 2013
  
\bibitem{Sally}S. Dawson and E. Furlan. A Higgs Conundrum with Vector
Fermions. Phys.Rev., D86:015021, 2012, arXiv:1205.4733.

\bibitem{atlas-bound}
 [ATLAS Collaboration], Note ATLAS-CONF-2013-018, [http://cds.cern.ch/record/1525525].


\bibitem{Carmi:2012yp}
  D.~Carmi, A.~Falkowski, E.~Kuflik and T.~Volansky,
  %``Interpreting LHC Higgs Results from Natural New Physics Perspective,''
  JHEP {\bf 1207} (2012) 136
  [arXiv:1202.3144 [hep-ph]].
  %%CITATION = ARXIV:1202.3144;%%
  %120 citations counted in INSPIRE as of 11 Apr 2013


\bibitem{Garberson:2013jz} 
  F.~Garberson and T.~Golling,
  %``Generalization of exotic quark searches,''
  arXiv:1301.4454 [hep-ex].
  
  \bibitem{Aad:2012bt} 
  G.~Aad {\it et al.}  [ATLAS Collaboration],
  %``Search for pair-produced heavy quarks decaying to Wq in the two-lepton channel at $\sqrt{s}=7$ TeV with the ATLAS detector,''
  Phys.\ Rev.\ D {\bf 86}, 012007 (2012)
  [arXiv:1202.3389 [hep-ex]].
  %%CITATION = ARXIV:1202.3389;%%
  %26 citations counted in INSPIRE as of 10 Apr 2013

\bibitem{Bonne:2012im} 
  N.~Bonne and G.~Moreau,
  %``Reproducing the Higgs boson data with vector-like quarks,''
  Phys.\ Lett.\ B {\bf 717}, 409 (2012)
  [arXiv:1206.3360 [hep-ph]].
  %%CITATION = ARXIV:1206.3360;%%
  %34 citations counted in INSPIRE as of 02 May 2013

\bibitem{Azatov:2012rj} 
  A.~Azatov, O.~Bondu, A.~Falkowski, M.~Felcini, S.~Gascon-Shotkin, D.~K.~Ghosh, G.~Moreau and S.~Sekmen,
  %``Higgs boson production via vector-like top-partner decays: Diphoton or multilepton plus multijets channels at the LHC,''
  Phys.\ Rev.\ D {\bf 85}, 115022 (2012)
  [arXiv:1204.0455 [hep-ph]].
  %%CITATION = ARXIV:1204.0455;%%
  %18 citations counted in INSPIRE as of 11 Apr 2013
  
\bibitem{CMS-PAS-TOP-12-035} [CMS Collaboration], Note CMS-PAS-TOP-12-035, [http://cds.cern.ch/record/1520879].                                                               

\bibitem{Kidonakis:2011wy} 
  N.~Kidonakis,
  %``Next-to-next-to-leading-order collinear and soft gluon corrections for t-channel single top quark production,''
  Phys.\ Rev.\ D {\bf 83}, 091503 (2011)
  [arXiv:1103.2792 [hep-ph]].
  %%CITATION = ARXIV:1103.2792;%%
  %141 citations counted in INSPIRE as of 11 Apr 2013
  
\bibitem{Aad:2012ux} 
  G.~Aad {\it et al.}  [ATLAS Collaboration],
  %``Measurement of the $t$-channel single top-quark production cross section in $pp$ collisions at $\sqrt{s}=7$ TeV with the ATLAS detector,''
  Phys.\ Lett.\ B {\bf 717}, 330 (2012)
  [arXiv:1205.3130 [hep-ex]].
  %%CITATION = ARXIV:1205.3130;%%
  %45 citations counted in INSPIRE as of 10 Apr 2013

\bibitem{Chatrchyan:2012ep} 
  S.~Chatrchyan {\it et al.}  [CMS Collaboration],
  %``Measurement of the single-top-quark $t$-channel cross section in $pp$ collisions at $\sqrt{s}=7$ TeV,''
  JHEP {\bf 1212}, 035 (2012)
  [arXiv:1209.4533 [hep-ex]].
  %%CITATION = ARXIV:1209.4533;%%
  %15 citations counted in INSPIRE as of 10 Apr 2013

\bibitem{Aad:2012ij} 
  G.~Aad {\it et al.}  [ATLAS Collaboration],
  %``A search for flavour changing neutral currents in top-quark decays in $pp$ collision data collected with the ATLAS detector at $\sqrt{s}=7$ TeV,''
  JHEP {\bf 1209}, 139 (2012)
  [arXiv:1206.0257 [hep-ex]].
  %%CITATION = ARXIV:1206.0257;%%
  %11 citations counted in INSPIRE as of 10 Apr 2013

\bibitem{AguilarSaavedra:2008gt} 
  J.~A.~Aguilar-Saavedra,
  %``Single top quark production at LHC with anomalous Wtb couplings,''
  Nucl.\ Phys.\ B {\bf 804}, 160 (2008)
  [arXiv:0803.3810 [hep-ph]].
  %%CITATION = ARXIV:0803.3810;%%
  %63 citations counted in INSPIRE as of 10 Apr 2013
  
  \bibitem{Fischer:2000kx} 
  M.~Fischer, S.~Groote, J.~G.~Korner and M.~C.~Mauser,
  %``Longitudinal, transverse plus and transverse minus $W$ bosons in unpolarized top quark decays at O(alpha($s$) ),''
  Phys.\ Rev.\ D {\bf 63}, 031501 (2001)
  [hep-ph/0011075].
  %%CITATION = HEP-PH/0011075;%%
  %58 citations counted in INSPIRE as of 10 Apr 2013
  
  \bibitem{Drobnak:2010ej} 
  J.~Drobnak, S.~Fajfer and J.~F.~Kamenik,
  %``New physics in $t-> b W$ decay at next-to-leading order in QCD,''
  Phys.\ Rev.\ D {\bf 82}, 114008 (2010)
  [arXiv:1010.2402 [hep-ph]].
  %%CITATION = ARXIV:1010.2402;%%
  %13 citations counted in INSPIRE as of 10 Apr 2013
  
  \bibitem{Aad:2012ky} 
  G.~Aad {\it et al.}  [ATLAS Collaboration],
  %``Measurement of the W boson polarization in top quark decays with the ATLAS detector,''
  JHEP {\bf 1206}, 088 (2012)
  [arXiv:1205.2484 [hep-ex]].
  %%CITATION = ARXIV:1205.2484;%%
  %25 citations counted in INSPIRE as of 10 Apr 2013
  
  \bibitem{Isidori:2003ts} 
  G.~D'Ambrosio, G.~Isidori and J.~Portoles,
  %``Can we extract short distance information from B(K(L) ---> mu+ mu-)?,''
  Phys.\ Lett.\ B {\bf 423}, 385 (1998)
  [hep-ph/9708326];
  %%CITATION = HEP-PH/9708326;%%
  %107 citations counted in INSPIRE as of 10 Apr 2013
  G.~Isidori and R.~Unterdorfer,
  %``On the short distance constraints from K(L,S) ---> mu+ mu-,''
  JHEP {\bf 0401}, 009 (2004)
  [hep-ph/0311084].
  %%CITATION = HEP-PH/0311084;%%
  %50 citations counted in INSPIRE as of 10 Apr 2013
   
  \bibitem{Inami:1980fz} 
  T.~Inami and C.~S.~Lim,
  %``Effects of Superheavy Quarks and Leptons in Low-Energy Weak Processes k(L) ---> mu anti-mu, K+ ---> pi+ Neutrino anti-neutrino and K0 <---> anti-K0,''
  Prog.\ Theor.\ Phys.\  {\bf 65}, 297 (1981)
  [Erratum-ibid.\  {\bf 65}, 1772 (1981)].
  %%CITATION = PTPKA,65,297;%%
  %1388 citations counted in INSPIRE as of 10 Apr 2013
  
  \bibitem{Buras:2012ru} 
  A.~J.~Buras, J.~Girrbach, D.~Guadagnoli and G.~Isidori,
  %``On the Standard Model prediction for BR(B{s,d} to mu+ mu-),''
  Eur.\ Phys.\ J.\ C {\bf 72}, 2172 (2012)
  [arXiv:1208.0934 [hep-ph]].
  %%CITATION = ARXIV:1208.0934;%%
  %55 citations counted in INSPIRE as of 10 Apr 2013
    
  \bibitem{Laiho:2009eu} 
  J.~Laiho, E.~Lunghi and R.~S.~Van de Water,
  %``Lattice QCD inputs to the CKM unitarity triangle analysis,''
  Phys.\ Rev.\ D {\bf 81}, 034503 (2010)
  [arXiv:0910.2928 [hep-ph]]. Updates available on http://latticeaverages.org/.
  %%CITATION = ARXIV:0910.2928;%%
  %153 citations counted in INSPIRE as of 10 Apr 2013
  
    \bibitem{Bouchard:2011xj} 
  C.~M.~Bouchard, E.~D.~Freeland, C.~Bernard, A.~X.~El-Khadra, E.~Gamiz, A.~S.~Kronfeld, J.~Laiho and R.~S.~Van de Water,
  %``Neutral $B$ mixing from $2+1$ flavor lattice-QCD: the Standard Model and beyond,''
  PoS LATTICE {\bf 2011}, 274 (2011)
  [arXiv:1112.5642 [hep-lat]].
  %%CITATION = ARXIV:1112.5642;%%
  %15 citations counted in INSPIRE as of 10 Apr 2013


  \bibitem{Lenz:2010gu} 
  A.~Lenz, U.~Nierste, J.~Charles, S.~Descotes-Genon, A.~Jantsch, C.~Kaufhold, H.~Lacker and S.~Monteil {\it et al.},
  %``Anatomy of New Physics in $B - \bar{B}$ mixing,''
  Phys.\ Rev.\ D {\bf 83}, 036004 (2011)
  [arXiv:1008.1593 [hep-ph]].
  %%CITATION = ARXIV:1008.1593;%%
  %152 citations counted in INSPIRE as of 10 Apr 2013
  
  \bibitem{Crivellin:2011ba} 
  A.~Crivellin and L.~Mercolli,
  %``$B -> X_d \gamma$ and constraints on new physics,''
  Phys.\ Rev.\ D {\bf 84}, 114005 (2011)
  [arXiv:1106.5499 [hep-ph]].
  %%CITATION = ARXIV:1106.5499;%%
  %16 citations counted in INSPIRE as of 10 Apr 2013
  
  \bibitem{delAmoSanchez:2010ae} 
  P.~del Amo Sanchez {\it et al.}  [BABAR Collaboration],
  %``Study of B ---> Xgamma decays and determination of |V_{td}/V_{ts}|,''
  Phys.\ Rev.\ D {\bf 82}, 051101 (2010)
  [arXiv:1005.4087 [hep-ex]];
  %%CITATION = ARXIV:1005.4087;%%
  %22 citations counted in INSPIRE as of 10 Apr 2013
  W.~Wang,
  %``$b -> s \gamma$ and $b -> d \gamma$ (B factories),''
  arXiv:1102.1925 [hep-ex].
  %%CITATION = ARXIV:1102.1925;%%
  %7 citations counted in INSPIRE as of 10 Apr 2013
  
  \bibitem{Aaij:2012nna} 
  RAaij {\it et al.}  [LHCb Collaboration],
  %``First evidence for the decay Bs -> mu+ mu-,''
  Phys.\ Rev.\ Lett.\  {\bf 110}, 021801 (2013)
  [arXiv:1211.2674 [Unknown]].
  %%CITATION = ARXIV:1211.2674;%%
  %92 citations counted in INSPIRE as of 10 Apr 2013
  
  
  \bibitem{Branco:1999fs} 
  G.~C.~Branco, L.~Lavoura and J.~P.~Silva,
  %``CP Violation,''
  Int.\ Ser.\ Monogr.\ Phys.\  {\bf 103}, 1 (1999).
  %%CITATION = IMPHA,103,1;%%
  %43 citations counted in INSPIRE as of 10 Apr 2013
  
    \bibitem{Haisch:2011up} 
  U.~Haisch and S.~Westhoff,
  %``Massive Color-Octet Bosons: Bounds on Effects in Top-Quark Pair Production,''
  JHEP {\bf 1108}, 088 (2011)
  [arXiv:1106.0529 [hep-ph]].


  \bibitem{Gresham:2012wc} 
  M.~I.~Gresham, I.~-W.~Kim, S.~Tulin and K.~M.~Zurek,
  %``Confronting Top AFB with Parity Violation Constraints,''
  Phys.\ Rev.\ D {\bf 86}, 034029 (2012)
  [arXiv:1203.1320 [hep-ph]].

    \bibitem{Dzuba:2012kx} 
  V.~A.~Dzuba, J.~C.~Berengut, V.~V.~Flambaum and B.~Roberts,
  %``Revisiting parity non-conservation in cesium,''
  arXiv:1207.5864 [hep-ph].

\bibitem{ALEPH:2005ab} 
  S.~Schael {\it et al.}  [ALEPH and DELPHI and L3 and OPAL and SLD and LEP Electroweak Working Group and SLD Electroweak Group and SLD Heavy Flavour Group Collaborations],
  %``Precision electroweak measurements on the $Z$ resonance,''
  Phys.\ Rept.\  {\bf 427}, 257 (2006)
  [hep-ex/0509008].
  %%CITATION = HEP-EX/0509008;%%
  %773 citations counted in INSPIRE as of 10 Apr 2013
  
\bibitem{Batell:2012ca} 
M.~Baak, M.~Goebel, J.~Haller, A.~Hoecker, D.~Kennedy, R.~Kogler, K.~Moenig and M.~Schott {\it et al.},
  %``The Electroweak Fit of the Standard Model after the Discovery of a New Boson at the LHC,''
  Eur.\ Phys.\ J.\ C {\bf 72} (2012) 2205
  [arXiv:1209.2716 [hep-ph]].
  %%CITATION = ARXIV:1209.2716;%%
  %37 citations counted in INSPIRE as of 10 Apr 2013
%  B.~Batell, S.~Gori and L.~-T.~Wang,
  %``Higgs Couplings and Precision Electroweak Data,''
  %JHEP {\bf 1301}, 139 (2013)
  %[arXiv:1209.6382 [hep-ph]].

  \bibitem{Bamert}
  P. Bamert, C.P. Burgess, James M. Cline, David London, and E. Nardi. R(b) and new physics:
A Comprehensive analysis. Phys.Rev., D54:4275�4300, 1996.


\bibitem{Lavoura:1992np} 
  L.~Lavoura and J.~P.~Silva,
  %``The Oblique corrections from vector - like singlet and doublet quarks,''
  Phys.\ Rev.\ D {\bf 47}, 2046 (1993).
  %%CITATION = PHRVA,D47,2046;%%
  %75 citations counted in INSPIRE as of 11 Apr 2013


\end{thebibliography}
\end{document}